\RequirePackage{fix-cm}
\documentclass[twocolumn]{svjour3}          
\smartqed  
\usepackage{graphicx}
\usepackage{amsmath}
\usepackage{amsfonts}
\usepackage{amssymb}
\usepackage[titletoc,title]{appendix}
\usepackage{color}
\usepackage{float}
\usepackage{caption}
\usepackage{subfig}

\newcommand{\sref}[1]{Sec.~\ref{#1}}
\newcommand{\aref}[1]{Appendix~\ref{#1}}
\newcommand{\eref}[1]{Eq.~(\ref{#1})}

\begin{document}
\title{Comparing Wigner, Husimi and Bohmian distributions: Which one is a true probability distribution in phase space?}

\author{E. Colom\'{e}s, Z. Zhan  and  X. Oriols }
\institute{E. Colom\'{e}s, Z. Zhan and  X. Oriols \at
              Departament d\rq{}Enginyeria Electr\`{o}nica, Universitat Aut\`{o}noma de Barcelona, Spain. \\
              \email{xavier.oriols@uab.es}          
           }

\date{\today}

\maketitle
\begin{abstract}

The Wigner distribution function is  a quasi-probability distribution. When  properly integrated, it provides the correct charge and current densities, but it gives negative probabilities in some points and regions  of the phase space. Alternatively, the Husimi distribution function is positive-defined everywhere, but it does not provide the correct charge and current densities. The origin of all these difficulties is the attempt to construct a phase space within a quantum theory that does not allow well-defined (i.e. simultaneous) values of the position and momentum of an electron.  In contrast, within the (de Broglie-Bohm) Bohmian theory of quantum mechanics, an electron has well-defined position and momentum. Therefore, such theory provides a \emph{natural} definition of the phase space probability distribution and by construction, it is positive-defined and it exactly reproduces the charge and current densities. The Bohmian distribution function has many potentialities for quantum problems, in general, and for quantum transport, in particular, that remains unexplored. 

\end{abstract}
\keywords{Wigner distribution \and  Husimi distribution \and Bohmian distribution \and Phase space}

\section{Introduction}

With our perspective of more than a century, the different attempts to construct phase space distributions for quantum phenomena seem a bit surprising. From the very beginning, after the work of de Broglie in 1926 \cite{broglie}, the big fathers of the quantum theory knew a formulation of quantum phenomena that leads to a true phase space distribution. According to de Broglie\rq{}s theory, it is possible to define simultaneously a position and a momentum for an electron. However, they preferred conscientiously to renounce this \emph{natural} path  for constructing \emph{true} phase space probability distributions. They looked for alternative ways to build a phase space distribution within the orthodox interpretation. All \emph{orthodox} attempts have inherent difficulties. For example, the Wigner or Husimi distribution functions are quasi-probability distributions. The prefix \emph{quasi} accounts for such difficulties. In this paper, we argue that de Broglie showed the path for a true phase space quantum distribution.

\subsection{Historical debates}

The beginning of the twentieth century brought surprising non-classical phenomena: Max Planck's explanation of the black body radiation \cite{omi.Planck-BlackBody}, the work of Albert Einstein on the photoelectric effect \cite{omi.Einstein-Photoelectric}, the Niels Bohr's model to account for the electron orbits around the nuclei \cite{omi.bohr}, etc. In order to provide an explanation for the underlying physics of such new phenomena, physicists were forced to abandon classical mechanics to develop novel, abstract and imaginative theories and formalisms.\\

In 1924, Louis de Broglie suggested in his doctoral thesis that matter, apart from its intrinsic particle-like behavior, could exhibit also a wave-like one \cite{omi.dB_AnnPhys}. Three years later he proposed an interpretation of quantum phenomena based on non-classical trajectories guided by a \emph{wave field}~\cite{omi.debroglie1927b}.
This was the origin of the pilot-wave formulation of quantum mechanics that we will refer as Bohmian mechanics to account for the following work of David Bohm \cite{omi.bohm1952a,omi.bohm1952b}. In the Bohmian formulation, an individual quantum system is formed by a point-like particle whose local velocity is defined from a guiding wave. The concept of an electron trajectory is, therefore, intrinsic to the Bohmian theory \cite{omi.Bohm1993,OriolsBook} and a phase space with well-defined position and momentum becomes a  \emph{natural} construction. 

Contemporaneously, Max Born and Werner Heisenberg, in the course of their collaboration in Copenhagen with Niels Bohr, provided an original explanation of all quantum phenomena without the need of trajectories \cite{omi.Born1926,omi.Heisenber1925}. This was the origin of the so-called Copenhagen interpretation (also known as the \emph{orthodox} formulation) of quantum phenomena. In the Copenhagen interpretation, an individual quantum system exhibits its wave or its particle nature depending on the experimental arrangement. However, it is not possible to retain the concept of trajectory because the theory itself forbids simultaneous definition of position and local velocity for an electron. Therefore, strictly speaking, there is no phase-space in the \emph{orthodox} formulation of quantum mechanics.    

In spite of these \emph{orthodox} difficulties, the scientific community has done a constant effort to effectively develop additional formalisms or theories\footnote{It is obvious that new mathematical elements open the path for different conceptual interpretations. The authors\rq{} point of view is that Bohmian mechanics and orthodox quantum mechanics are indeed two different theories, while the Wigner distribution function is a new formalism belonging to the orthodox school. In any case, such distinction is not clear at all and it is not relevant for the conclusions of this work.} to define a quantum phase space with position $x$ and momentum $p$. The first attempt was due to  Hermann Weyl \cite{weyl} in 1927 and Eugene Wigner \cite{omi.wigner} in 1932, who developed the so called Wigner-Weyl  transform between the Hilbert space and a new quantum phase space. The independent works realized by Hip Groenewold \cite{groenewold} and Joe Moyal \cite{moyal} lead to a complete formalism where the time evolution of a quantum system itself was defined in phase space alone (not in the Hilbert space). For a proper discussion see also \cite{bayen1,bayen2,baker,Zachos}. The Wigner function is a quasi-probability distribution because it gives negative values at some regions of the phase space.  Of course, such negative values do not invalidate the physical contents of the Wigner function because it reproduces the correct charge and current densities when properly integrated. In addition, some relevant works do also give physical explanation to these negative values \cite{Dirac,Monroe,feynman,review}. Their main argument is that as far as we cannot measure these probabilities, they are a correct and acceptable mathematical tool. At the end of the day, only positive probabilities corresponding to measurable observables will be recorded. In 1940, Kodi Husimi\cite{omi.husimi} developed a non-negative phase-space probability function, the so-called Husimi distribution. However, in spite of being a non-negative phase-space distribution, it is also a quasi-probability distribution because it does not reproduce correctly values of the charge and current densities when integrated. 
   
By construction, Bohmian mechanics, orthodox quantum mechanics and Wigner distribution functions give exactly the same empirical results for all (non-relativistic) quantum mechanical phenomena. Therefore, all of them are perfectly valid tools to develop practical solutions for quantum problems. Every formalism has pros and cons. For example, when dealing with phase space,  it is unquestionable that the Wigner distribution has had a large practical utility in many quantum problems, even in our days \cite{kerry,frensley,sellier,weinbub}. This special issue devoted to the use of the Wigner function in the computation of quantum electronics shows its unquestionable success. Therefore, the main goal of this paper is not to criticize the quasi-probability nature of the Wigner distribution, but to point out that Bohmian mechanics provides an original quantum theory with well-defined phase space. It is possible to construct a \emph{true} distribution of probability in phase space. At the present moment, it is not evident if such Bohmian distributions can be as useful as the Wigner ones, but it seems obvious that this true phase space distribution merits to be investigated.

 
\subsection{Phase space in Classical mechanics}

In classical mechanics, a particle has well-defined position $x(t)$ and momentum $p(t)$. Then, if we consider a large number $N$ of trajectories (for example, by repeating the same experiment with different initial conditions), it is possible to define a classical phase space distribution $F_c(x,p,t)$ by counting the number of trajectories at each point $\{x,p\}$ of the phase space:

\begin{equation}
F_c(x,p,t)=\frac{1}{N}\sum_{i=1}^N\delta(x-x_i(t))\delta(p-p_i(t))
\label{distribution_classical}
\end{equation}
where $x_i(t)$ and $p_i(t)$ are the actual position and momentum of the $i$-th particle at time $t$ with $i=1,...,N$. The evolution of this classical phase space distribution can be found by directly solving the Newton laws of the $N$ trajectories or by solving the Boltzmann equation. \\

At this point, we want to clarify that the discussion done in this paper will be focused on one-particle systems. The function $F_c(x,p,t)$ is constructed by repeating the experiment $N \rightarrow \infty$ times (or by dealing simultaneously with $N$ independent particles). All conclusions elaborated in this paper about one-particle phase space probability functions can be straightforwardly generalized for many-particle systems, but with a large increment of notation complexity that would later become irrelevant for our conclusions. \footnote{Strickly speaking, the phase space of a system of 2 interacting particles is not $\{x,p\}$ but $\{x_1,x_2,p_1,p_2\}$. } For the same reason, we will consider only one spatial degree of freedom and one momentum degree of freedom for each particle.  

\subsection{Phase space in Quantum mechanics}

The proper definition of the \emph{phase space} is not so obvious in quantum mechanics. First, let us notice that there are several different but \emph{empirically} equivalent theories that successfully account for all quantum phenomena.  Strictly speaking, there is no phase space in the Copenhagen (\emph{orthodox}) quantum theory. The local position and local momentum do not exist, simultaneously, in that theory. 


Let us start by discussing why there is no phase space in the Copenhagen (\emph{orthodox}) quantum theory. According to Dirac, the observables are represented by operators in Hilbert space \cite{Dirac}. The commutator of these operators specifies which quantities can be known simultaneously and which not  \cite{Dirac} . In the case of the position operator $\hat{x}$ and the momentum operator $\hat{p}$, we have:

\begin{equation}
[\hat{x},\hat{p}]|\psi \rangle=\hat{x}\hat{p}|\psi\rangle-\hat{p}\hat{x}|\psi\rangle=i\hbar|\psi\rangle
\label{commutator}
\end{equation}

Therefore the orthodox theory does not support the simultaneous knowledge of local positions and momenta. Strictly speaking, there is no phase space in the orthodox interpretation of quantum mechanics. Nevertheless, within the orthodox quantum mechanics, we will see later that the Wigner formalism defines a phase space with a well-defined value for the position and the momentum of particles, through the use of the Wigner-Weyl transform.


\subsection{Definition of a probability distribution}

Let us specify what we understand by a well-defined probability distribution in the phase space, say $F_{Q}(x,p)$, for a quantum (or classical) system. One desires that this probability distribution fulfils the probabilities axioms\footnote{In the text, all the integration limits are from $-\infty$ to $\infty$ and thus we will not write them explicitly.}:

\begin{eqnarray}
&&F_{Q}(x,p) \geq 0 \label{cond1},\\
&&\int  \int  F_{Q}(x,p) dx dp = 1 \label{cond2}. 
\end{eqnarray}

In addition, its marginal distribution should give the usual position or charge probability distributions\footnote{For simplicity, we avoid the explicitly consideration of the charge $q$ of an electron in \eref{marg1} and \eref{current}. We notice also that the words charge or current densities can be misleading when a wave packet is partially transmitted or reflected, and the charge is in fact either fully transmitted or fully reflected, not both, when measured.}:

\begin{eqnarray}
 Q(x) =\int F_{Q}(x,p) dp ,
\label{marg1}
\end{eqnarray}
Expression (\ref{marg1}) is an important quantity in quantum transport because it is related to the charge density, which is a very relevant magnitude in any self-consistent solution of the electron transport. Another important quantity built from this distribution is the current density, which can be expressed as:

\begin{eqnarray}
J(x)=\int pF_{Q}(x,p)dp.
\label{current}
\end{eqnarray}

Several attempts have been done trying to provide such a phase space quantum distribution $F_{Q}(x,p)$. Hereafter we report the Wigner distribution function \cite{omi.wigner} (which is nothing else but a mathematical Wigner-Weyl transform of the density matrix) and the Husimi distribution (a smoothed version of the Wigner distribution function). The negative values of the Wigner distribution function in some regions of the phase space avoids the possibility of this Wigner distribution function to be a true probability function.  The Husimi distribution, by construction, has non-negative \cite{omi.husimi} values. However, it does not reproduce correctly the marginal distributions (\ref{marg1}) and (\ref{current}).\\

If one is interested in a \emph{true} probability distribution in phase space, it seems appropriate to use a quantum theory that has a well-defined phase space, i.e. a theory that explicitly accounts for well-defined positions and local momenta \cite{omi.bohm1952a,omi.bohm1952b}. Such theory exist. The Bohmian theory briefly mentioned in the first paragraph is the desired theory. \footnote{It must be said that, due to the measurement processes, position and momentum are not accessible simultaneously in a laboratory in a single experiment as seen in expression (\ref{commutator}), but the Bohmian theory supports the (ontological) definition of both quantities.}. In Bohmian mechanics, the electron is defined at any time by a position plus a wave-function. The wave-function provides the local velocity of the particle that provides the electron trajectory when properly integrated. In summary, Bohmian mechanics is a theory which provides a well-defined position and momentum for a particle. Thus, contrarily to the orthodox theory,  the Bohmian theory allows the existence of a \emph{physical} and \emph{natural} phase space. By construction, such Bohmian probability distribution is non-negative and it satisfies all the probability axioms in order to be a correct probability distribution. \emph{If the scientific community is interested in describing quantum phenomena in phase space, why not using the Bohmian distribution function ?}\\

After this large introduction, in \sref{sec2} Wigner, Husimi and Bohmian distributions are defined and we compare their properties. In \sref{sec3} we show numerical results for the phase space of the mentioned distributions when an electron impinges on a double barrier. We discuss and show numerically some disadvantages of the Wigner and Husimi distributions, which are not present in the Bohmian distribution.  Finally, in \sref{sec4} conclusions are exposed. Several appendixes discuss many technical details omitted in the main text.  

\section{Quantum phase space probability distribution}
\label{sec2}

Next, we define and compare the Wigner, Husimi and Bohmian phase space probability distributions for quantum systems. 

\subsection{Wigner distribution}

One quite common way of describing a quantum mechanical system in phase space is by the so-called Wigner distribution ($F_W$). For a given state $|\psi\rangle$, one can construct the  density matrix operator $\hat{\rho}=|\psi\rangle\langle  \psi|$ and express it in the position representation $\langle x|\hat{\rho}|x\rq{}\rangle\!\!=\!\!\langle  x|\psi\rangle\langle  \psi|x\rq{}\rangle$ or in the  momentum representation $\langle  p|\hat{\rho}|p\rq{}\rangle\!\!=\!\!\langle  p|\psi\rangle\langle  \psi|p\rq{}\rangle$. Therefore, somehow, the Wigner distribution can be interpreted as an intermediate representation between this two and it is given by a Wigner-Weyl transform of the density matrix:

\begin{eqnarray}
F_W(x,p) = \frac{1}{h} \int  \psi(x+\frac{y}{2}) \psi^{*}(x-\frac{y}{2})e^{i\frac{py}{\hbar}} dy.
\end{eqnarray}
where $\psi(x)=\langle  x|\psi\rangle$. We notice that the (mathematical) variable $p$ in the Wigner (quasi) phase space $\{x,p\}$ appears due to a Fourier transform of a type of autocorrelation function of the wave function. In the case of mixed states, the density matrix can be written as $\hat{\rho}=\sum_j c_j |\psi_j\rangle \langle  \psi_j|$ where $c_j$  specifies the fraction of the ensemble in the pure state $|\psi_j\rangle$. For the sake of simplicity we avoid the explicit time dependence of the wave function and $c_j$. Therefore, the Wigner distribution function for a mixed state is the following:

\begin{eqnarray}
\label{wigner-psi}
F_W(x,p) = \!\frac{1}{h} \sum_j c_j \!\! \int \! \psi_j(x+\frac{y}{2}) \psi_j^{*}(x-\frac{y}{2}) e^{i p y / \hbar} dy.
\end{eqnarray}

Hereafter, being the extension to mixed states straightforwardly \cite{Ballentine} achieved without any modification in the conclusions reached in this paper, for simplicity, we will consider only pure states.

The Wigner distribution is a quasi-probability distribution because it does not satisfy, in general, the condition given in \eref{cond1} for a well-defined phase space probability distribution. In \aref{neg_wig} we can see why this distribution can be negative at some regions of the phase space. In addition, in \sref{sec3} we report a numerical example where the probability in the phase space is clearly negative.

We can calculate also $Q_W(x)$ and $J_W(x)$ directly using \eref{marg1} and \eref{current}. The results obtained (the derivation can be seen at \aref{wigner_derivations}) are:

\begin{eqnarray}
Q_W(x) = |\psi(x)|^2
\label{Q_W}
\end{eqnarray}

\begin{eqnarray}
J_W(x) = |\psi(x)|^2 \frac{\partial S(x)}{\partial x} ,
\label{J_W}
\end{eqnarray}

being $S(x)$ the angle of the polar representation of the wave function $\psi(x,t)=R(x)exp(iS(x)/\hbar)$ (for a detailed explanation see  \aref{wigner_derivations}). The results (\ref{Q_W}) and (\ref{J_W}) are the ones that one will get by directly using $|\psi\rangle$.  At this point, because of (\ref{Q_W}) and (\ref{J_W}), we see that $F_W(x,p)$ is a good candidate to study quantum transport. However, we will see later that it is \lq\lq{}dangerous\rq\rq{} to take seriously the Wigner (quasi) phase space when further developing the basic steps described here. For example, when including transitions between the phase space points $\{x,p\}$ and $\{x,p\rq{}\}$ due to the (Fermi Golden rule) scattering. 

\subsection{Husimi distribution}

The Husimi distribution ($F_H$) is another possible phase space distribution built from the Copenhagen school. In this case, it does satisfy the condition (\ref{cond1}) by construction. For this purpose, we use a set of minimum non-orthogonal uncertainty states localized in phase space ($|q,p\rangle$) \cite{Ballentine}. Using them, the Husimi distribution is the following:

\begin{eqnarray}
F_H(x,p) \!\!=  \frac{1}{2\pi \hbar} \langle   x,p|\hat{\rho} |q,p\rangle.
\label{Husimi1}
\end{eqnarray}

If we use that $\hat{\rho}=|\psi  \rangle\langle   \psi |$, we achieve the following equation where it is reflected the positiveness of the Husimi distribution:

\begin{eqnarray}
F_H(x,p) \!\!= \!\!\frac{1}{2\pi \hbar}  \langle    x,p|\psi  \rangle\langle   \psi |x,p  \rangle= \!\!\frac{1}{2\pi \hbar} |\langle   x,p|\psi \rangle|^2.
\label{Husimi2}
\end{eqnarray}

However, it is not also a true probability distribution because it does not fulfil the marginal property (\ref{marg1}). It can be also seen that the Husimi distribution is just a Gaussian smoothed\footnote{We notice that the (mathematical) variables $x$ and $p$ in the Husimi (quasi) phase space $\{x,p\}$ are a smoothed version of the $x$ and $p$ variables of the Wigner (quasi) phase space.}  version of the Wigner distribution \cite{Ballentine}:

\begin{eqnarray}
F_H(x,p) \!\!= \!\!\frac{1}{\pi \hbar}\!\! \int \!\!\!F_W(x',p') e^{\frac{-(x-x')^2}{2s^2}}\!e^{\frac{-(p-p')^2 2s^2}{\hbar^2}}dx'\!dp'.
\label{Husimi}
\end{eqnarray}

Next, we will calculate the charge and current densities $Q_H(x)$ and $J_H(x)$ similarly as done for the Wigner function (to see the complete derivation see \aref{wigner_derivations}):

\begin{eqnarray}
Q_H(x) = \frac{1}{\sqrt{2\pi s^2}}\int|\psi(x')|^2 e^{-\frac{(x-x')^2}{2s^2}}dx',
\label{Q_H}
\end{eqnarray}

\begin{eqnarray}
J_H(x) = \frac{1}{\sqrt{(2\pi s^2)}}\int R^{2}(x')\frac{\partial S(x')}{\partial x'}e^{-\frac{(x-x')^2}{2s^2}}dx'.
\label{J_H}
\end{eqnarray}

We can clearly see, that these results are the ones obtained for the Wigner function, but smoothed by a Gaussian function. From \eref{Husimi} we can understand why the Husimi distribution does not accomplish \eref{marg1}. The broadening of the probabilities changes the momentum and position distributions. For these reasons it is considered also a quasi-probability distribution. The difficulties in properly providing the current and charge densities are a dramatic drawback for the correct simulation of quantum electronic devices with the Husimi distribution.

\subsection{Bohmian distribution}

In order to explain the Bohmian phase space distribution, we briefly explain Bohmian mechanics. It is a theory related to waves and particles. The evolution of the wave function is a solution of the typical Schr\"odinger equation:

\begin{equation}
i\hbar\frac{\partial \psi(x,t)}{\partial t}=H\psi(x,t).
\label{Scho}
\end{equation}

The wave function itself defines a local velocity for particles,

\begin{equation}
v(x,t) =  \frac{\hbar}{m}\text{Im} \frac{\nabla \psi}{\psi} = \frac{\hbar}{m}\frac{\partial S(x_1)}{\partial x_1}.
\label{vel}
\end{equation}

Let us emphasize that particles, here, mean point-like particles. For a more detailed explanation see \cite{OriolsBook}. With these two equations, particles can be described by trajectories which have, as it has been already commented, a definite position and momentum:

\begin{equation}
x(t)=x(0)+\int_0^t v(x,t)dt.
\label{traj}
\end{equation}

Once we have well-defined trajectories, we can again compute the quantum Bohmian phase space distribution\footnote{We notice that the  variables $x$ and $p$ in the Bohmian phase space $\{x,p\}$ are directly defined in the theory itself.  They are part of the ontology of Bohmian mechanics. For this reason, the Bohmian phase space is a \emph{natural} space, without mathematical tricks.} similarly to the classical case:

\begin{equation}
F_B(x,p,t)=\lim_{N \to \infty}\frac{1}{N}\sum_{i=1}^N\delta(x-x_i(t))\delta(p-p_i(t)),
\label{distribution_bohmian}
\end{equation}

where $N$ is the number of different trajectories of an ensemble of experiments, each experiment has a different initial position\footnote{Let us emphasize that the different $x_i(t)$ and $p_i(t)$ are not associated to different particles (as we have said, for simplicity, throughout the paper we only deal with single-particle one-degree-of-freedom  problems), but to different realizations of the same experiment. The probability obtained from the wave function $\psi(x,t)$ has exactly the same (ensemble) meaning.}. In \eref{distribution_bohmian}, $x_i(t)$ is a position of the trajectory in \eref{traj} at time $t$, while $p_i(t)=m v(x_i(t),t)$ is the momentum of the particle related to the velocity in \eref{vel} with the electron mass $m$. Let us emphasize that, by construction, the phase space distribution constructed with Bohmian mechanics is always non-negative. The number of Bohmian trajectories with momentum $p_B$ at the position $x$ must be positive (or zero if there are no particles).

At this point, we want to emphasize an important clarification about the experimental measurement of the Bohmian position and momentum. Because of expression $(\ref{commutator})$, in a single experiment, one can measure the position or the momentum, not both. However, contrarily to the Copenhagen school, the Bohmian theory does not renounce to a well-defined definition of position and momentum because of the mentioned experimental limitation. In addition, as we will further discuss in the conclusions, the Bohmian phase space distribution is experimentally accessible from a proper treatment of the experimental data obtained from  a  weak measurement of momentum and positions in a large ensemble of experiments. 

From the Bohmian distribution, we can calculate the charge and current densities.

\begin{eqnarray}
Q_B(x) = |\psi(x)|^2,
\label{Q_B}
\end{eqnarray}

\begin{eqnarray}
J_B(x) = |\psi(x)|^2 \frac{\partial S(x)}{\partial x}.
\label{J_B}
\end{eqnarray}

As we can see, these results are exactly the same as the ones obtained from the Wigner distribution (and different from the ones obtained from the Hussimi distribution). Therefore, the Bohmian distribution is an excellent tool to study quantum electron transport. In addition, in the next section we will show how the Bohmian phase space distribution does not have the problems found for the Wigner and the Husimi distributions.

\section{Numerical example}
\label{sec3}

According to the conceptual discussions in the previous sections, here, we provide numerical examples for the three mentioned quantum phase space distributions and the related charge density and current density. For simplicity, we consider a simple one-dimensional Gaussian wave packet impinging in a symmetric double  barrier. At the initial time $t_0$, the wave function of a Gaussian wave packet at the left of the barrier is :

\begin{equation}
\label{Gaussian}
\psi(x,t_0)=(\frac{1}{2\pi a_0^2})^{\frac{1}{4}}e^{i k_0(x-x_0)}exp \left( -\frac{(x-x_0)^2}{4 a_0^2} \right),
\end{equation}
where $a_0=7.5\:nm$ is the initial spatial variance of the wave packet, $x_0=100\:nm$ is the initial central position and $k_0=0.69\:nm^{-1}$ is the central wave vector. In addition, for the Husimi evolution, we used also the same dispersion: $s=7.5\:nm$.

\begin{figure}[h]
\begin{minipage}{18cm}
  \centering
      \subfloat[Wigner distribution at $t_0$.]{\label{wf0}
          \includegraphics[width=0.50\textwidth]{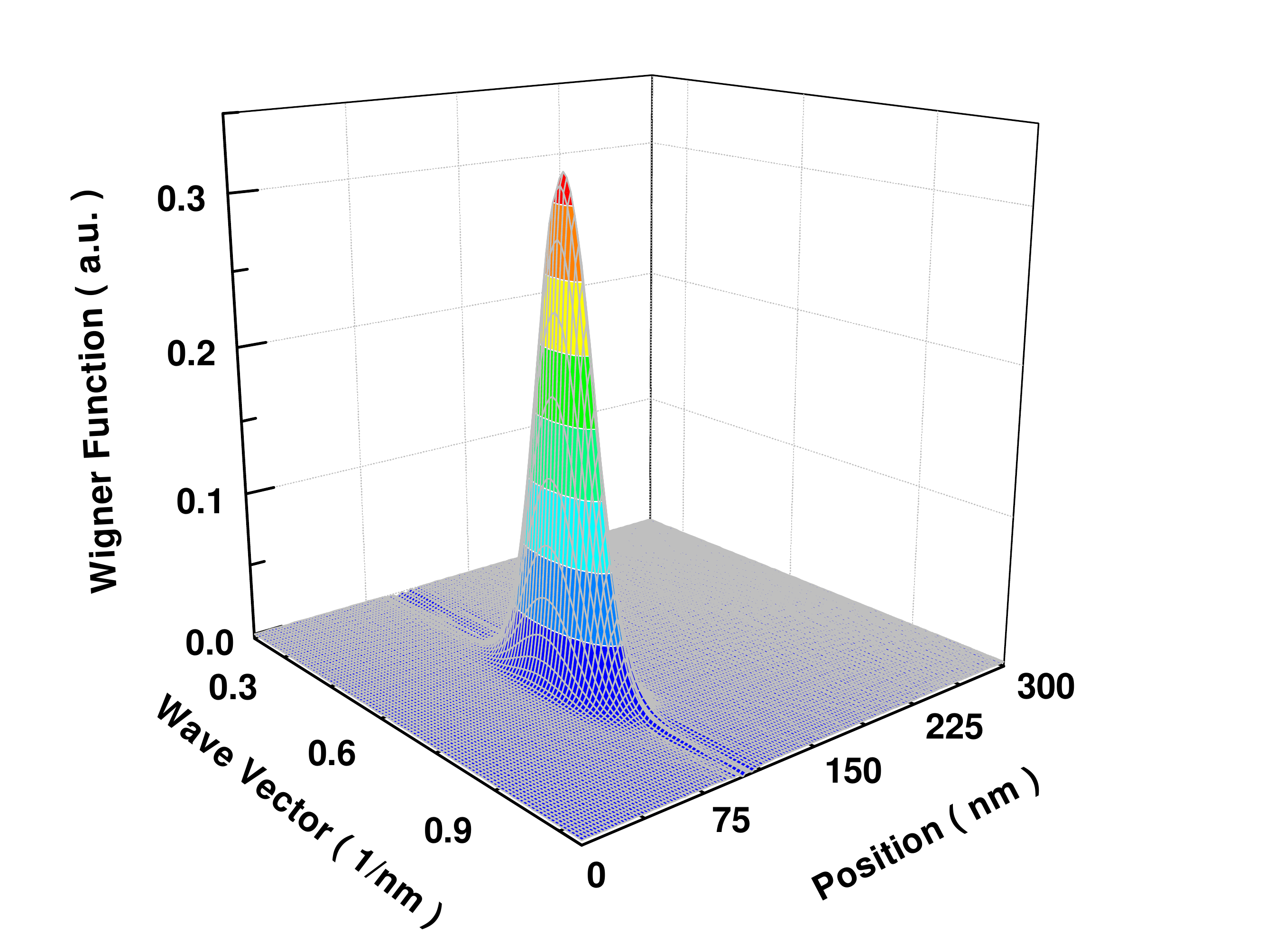}}
           \subfloat[Wave packet impinging on a tunneling barrier at $t_0$.]{\label{wp0}
          \includegraphics[width=0.50\textwidth]{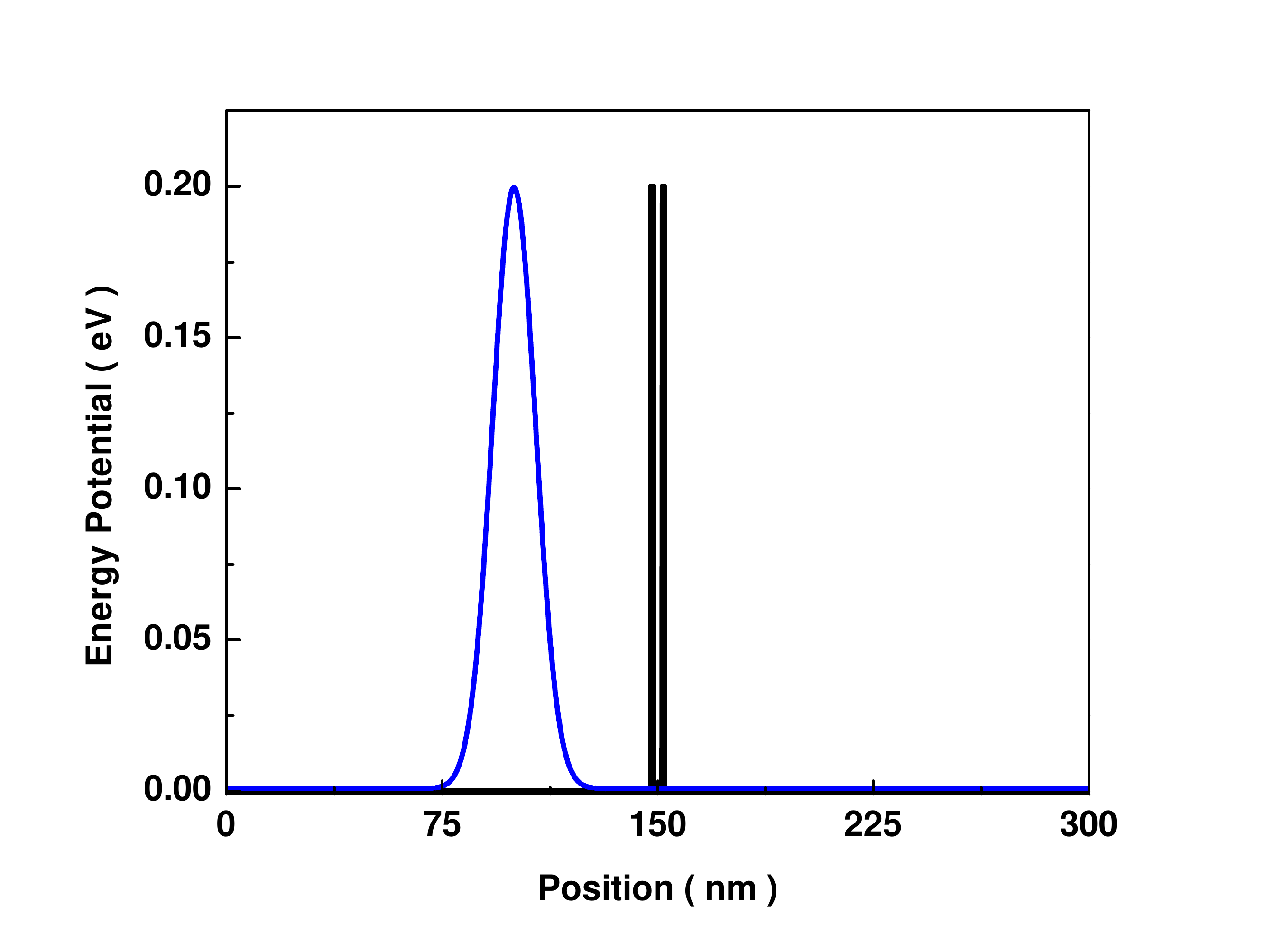}}\quad
          \subfloat[Husimi distribution at $t_0$.]{\label{hf0}
          \includegraphics[width=0.50\textwidth]{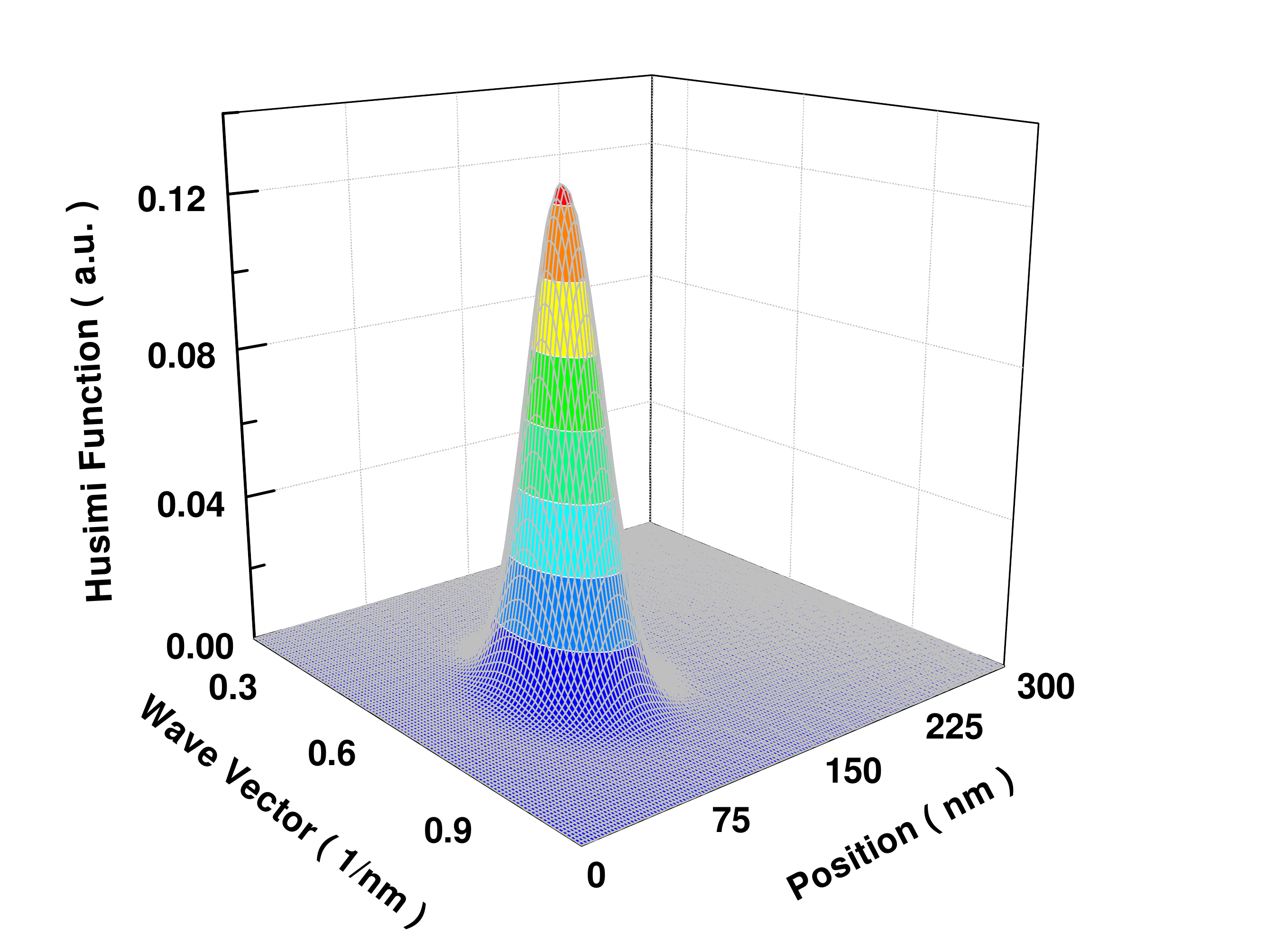}}
           \subfloat[Charge density for the three quantum phase space distributions at $t_0$.]{\label{ch0}
          \includegraphics[width=0.50\textwidth]{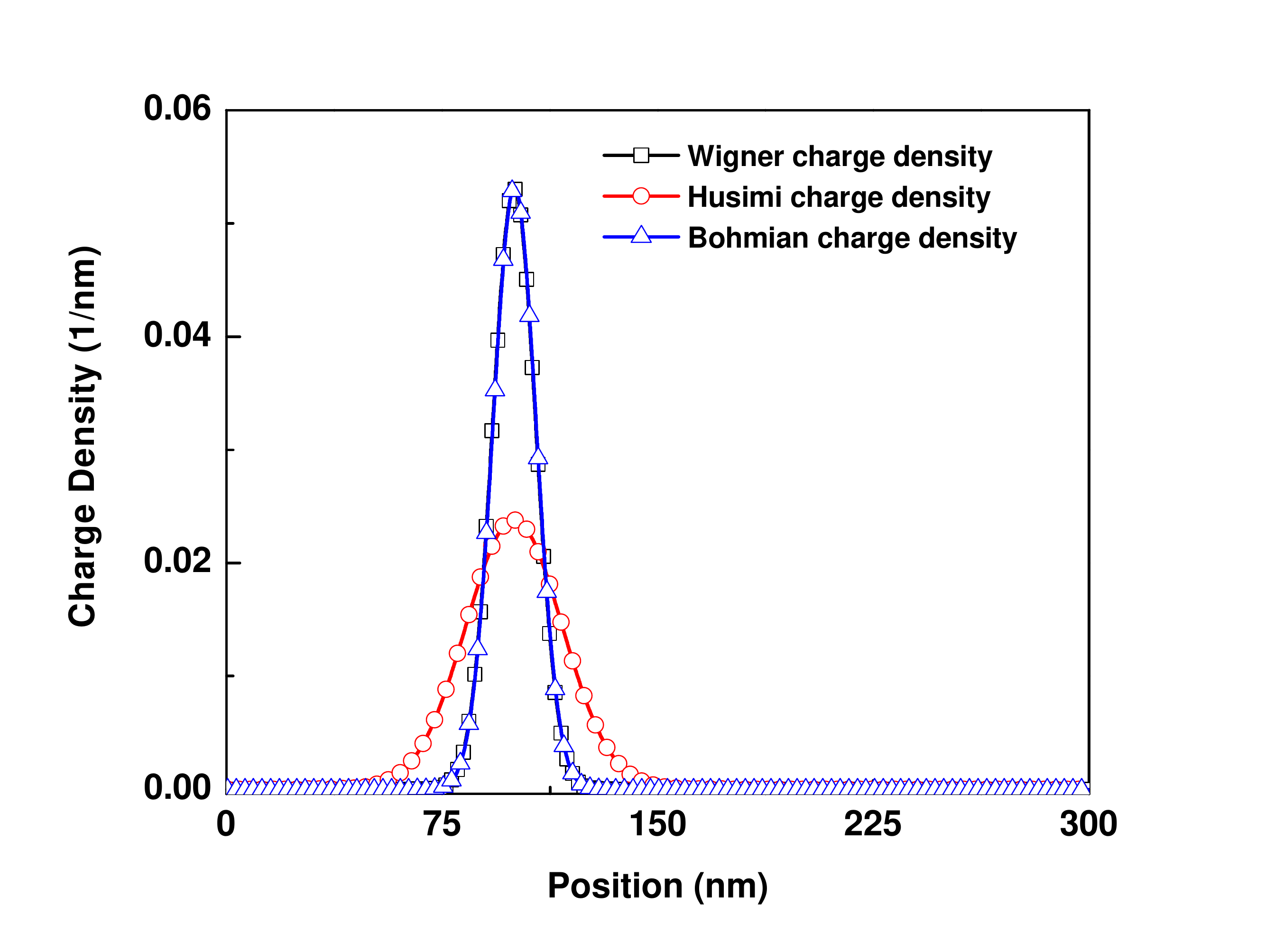}}\quad
          \subfloat[Bohmian distribution at $t_0$.]{\label{bf0}
          \includegraphics[width=0.50\textwidth]{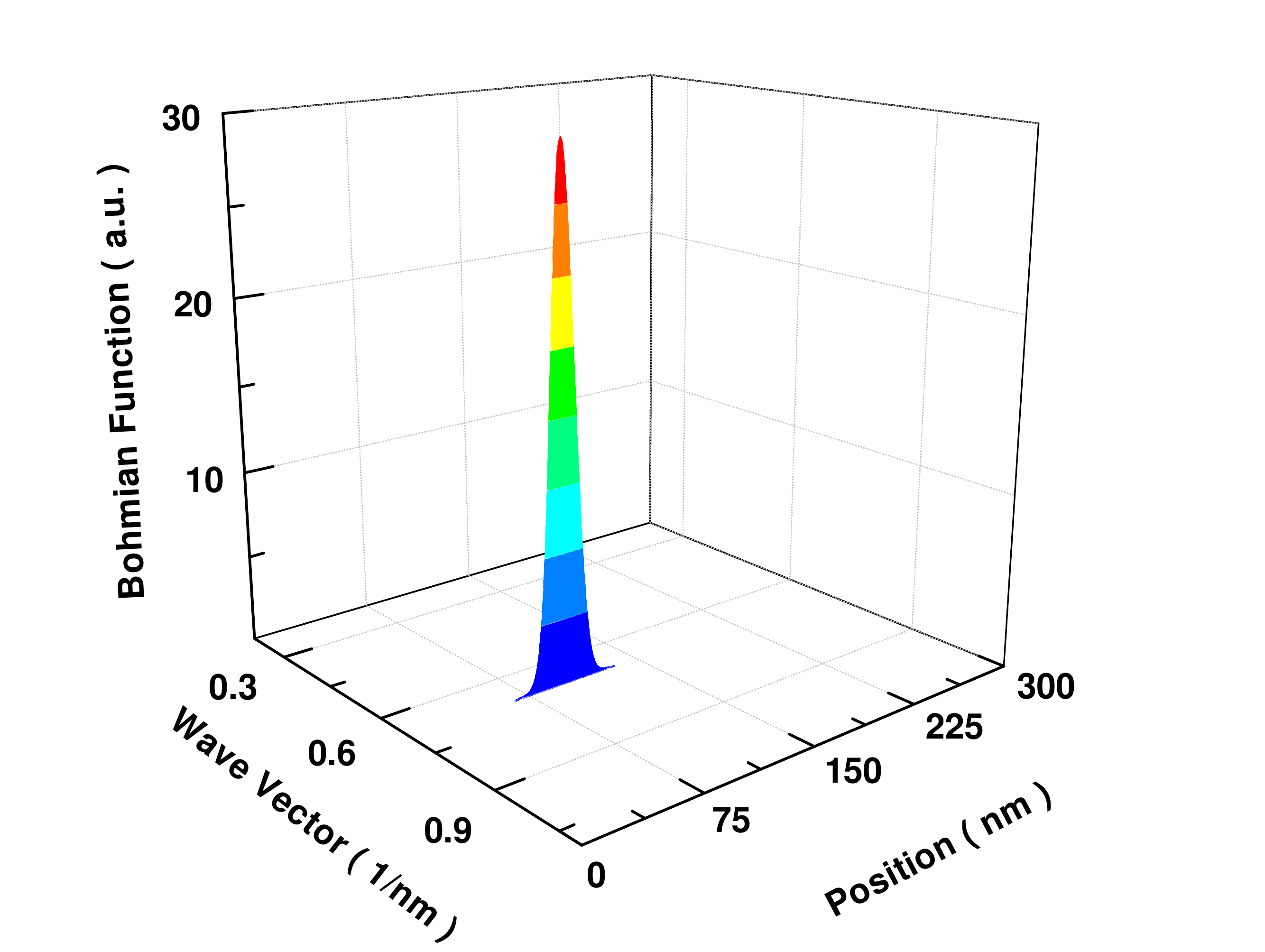}}
          \subfloat[Current density for the three quantum phase space distributions at $t_0$.]{\label{cu0}
          \includegraphics[width=0.50\textwidth]{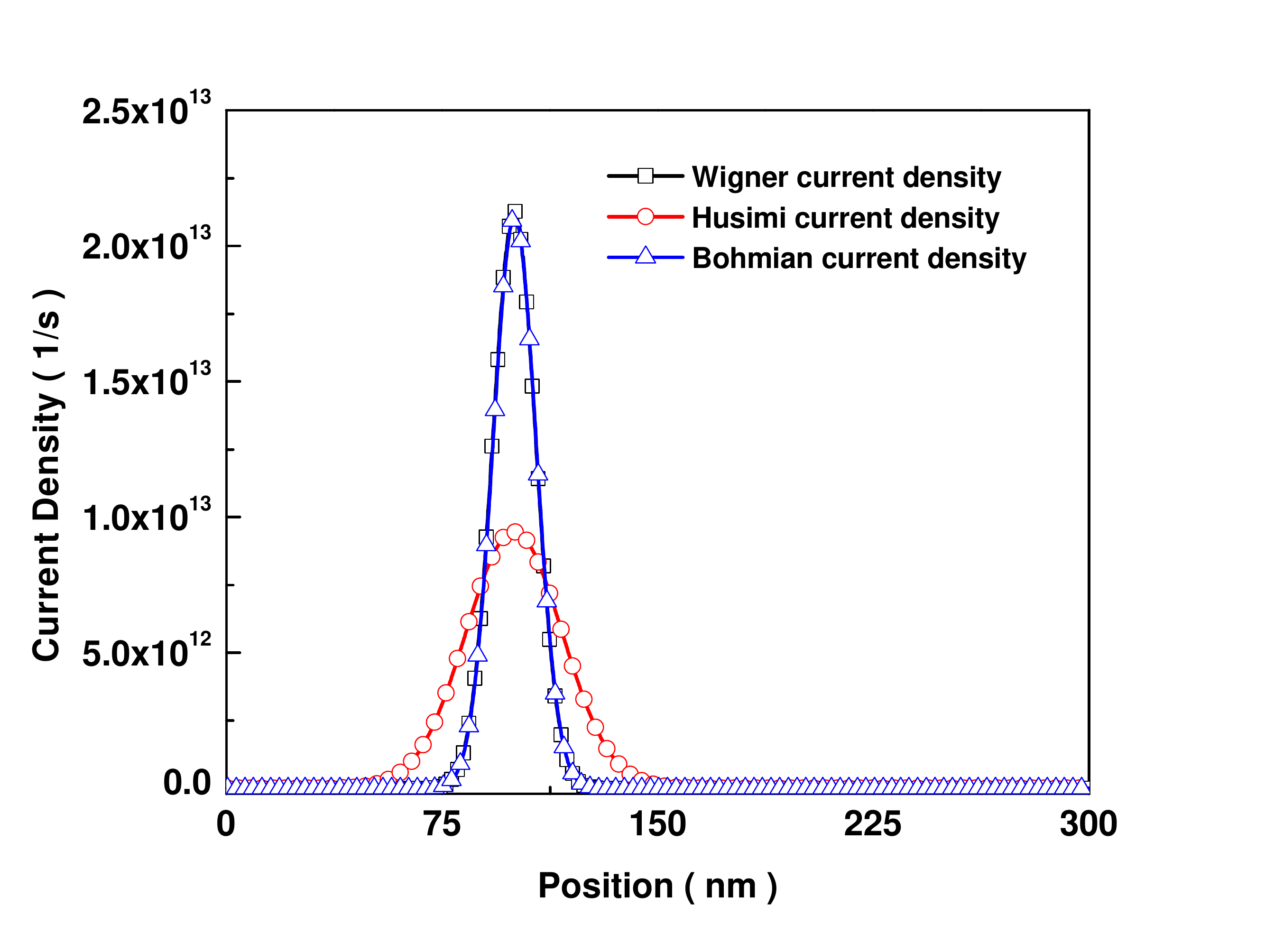}}
     \caption{Simulation of the (a) Wigner distribution, (c) Husimi distribution and (e) Bohmian distribution at the initial time $t_0$. (b) simulation of the wave packet impinging on a double barrier, the simulation parameters are: $E=0.09\:eV$, $m^*\!=\!0.2m_0$, where $m_0$ is the free-electron mass, the barrier height is $0.2\:eV$, the barrier width is $0.8\:nm$ and the well depth is $3.2\:nm$ (d) and (f) are the charge and current densities for the three phase space distributions, respectively.}
  \label{fig1}
 \end{minipage}
\end{figure}

\begin{figure}[h]
\begin{minipage}{18cm}
  \centering
      \subfloat[Wigner distribution at $t_1$.]{\label{wf1}
          \includegraphics[width=0.50\textwidth]{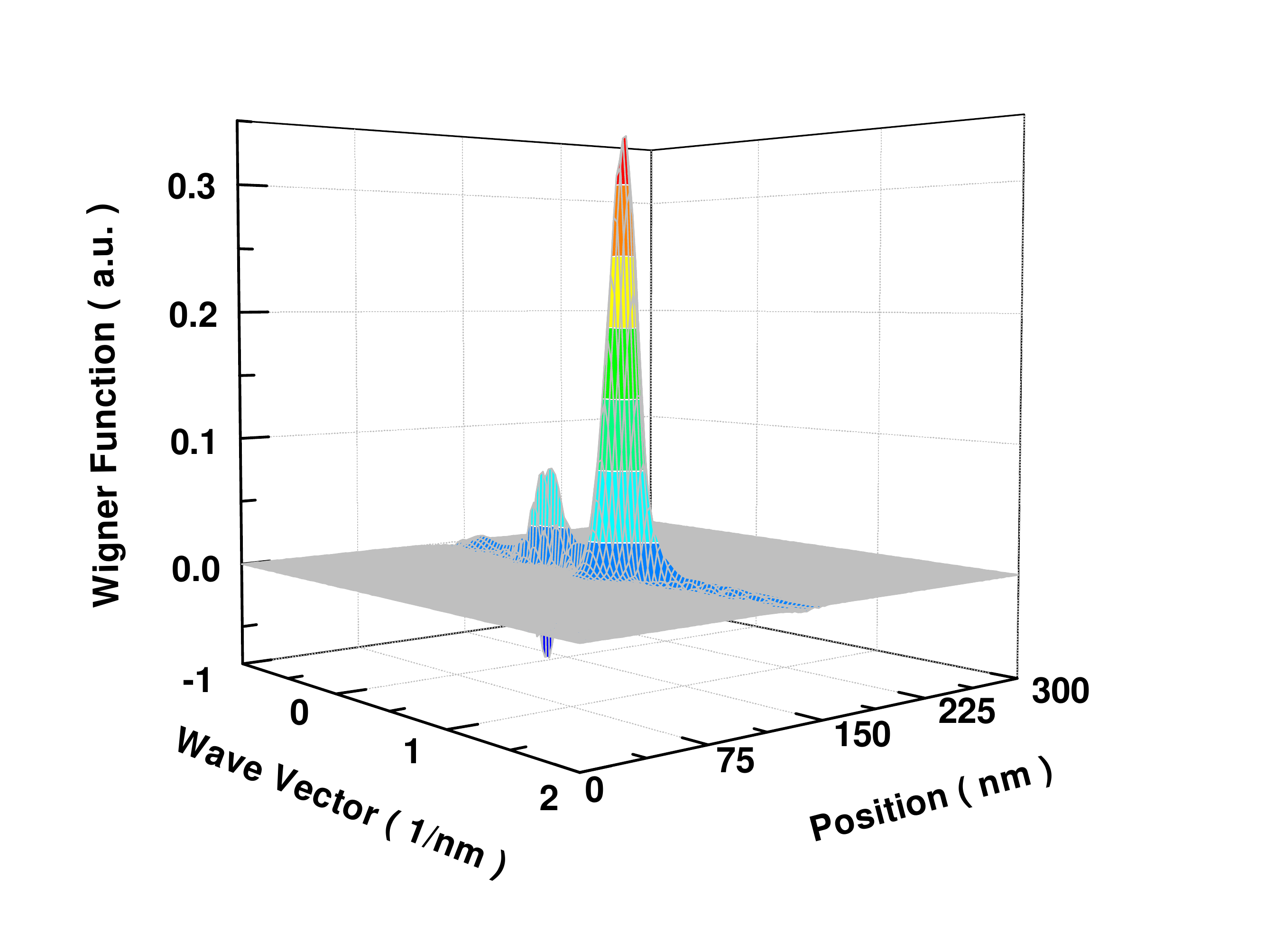}}
           \subfloat[Wave packet impinging on a tunneling barrier at $t_1$.]{\label{wp1}
          \includegraphics[width=0.50\textwidth]{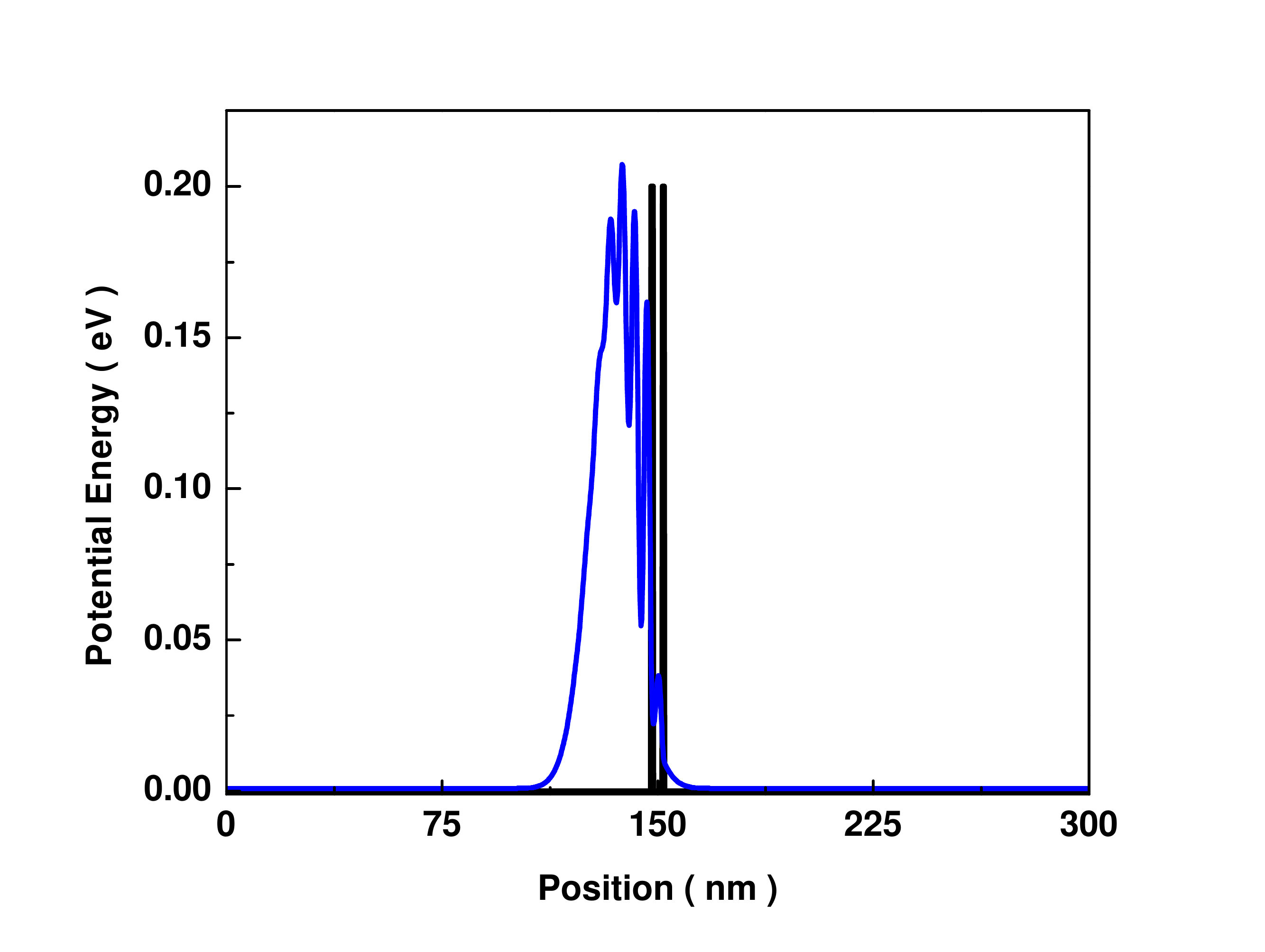}}\quad
          \subfloat[Husimi distribution at $t_1$.]{\label{hf1}
          \includegraphics[width=0.50\textwidth]{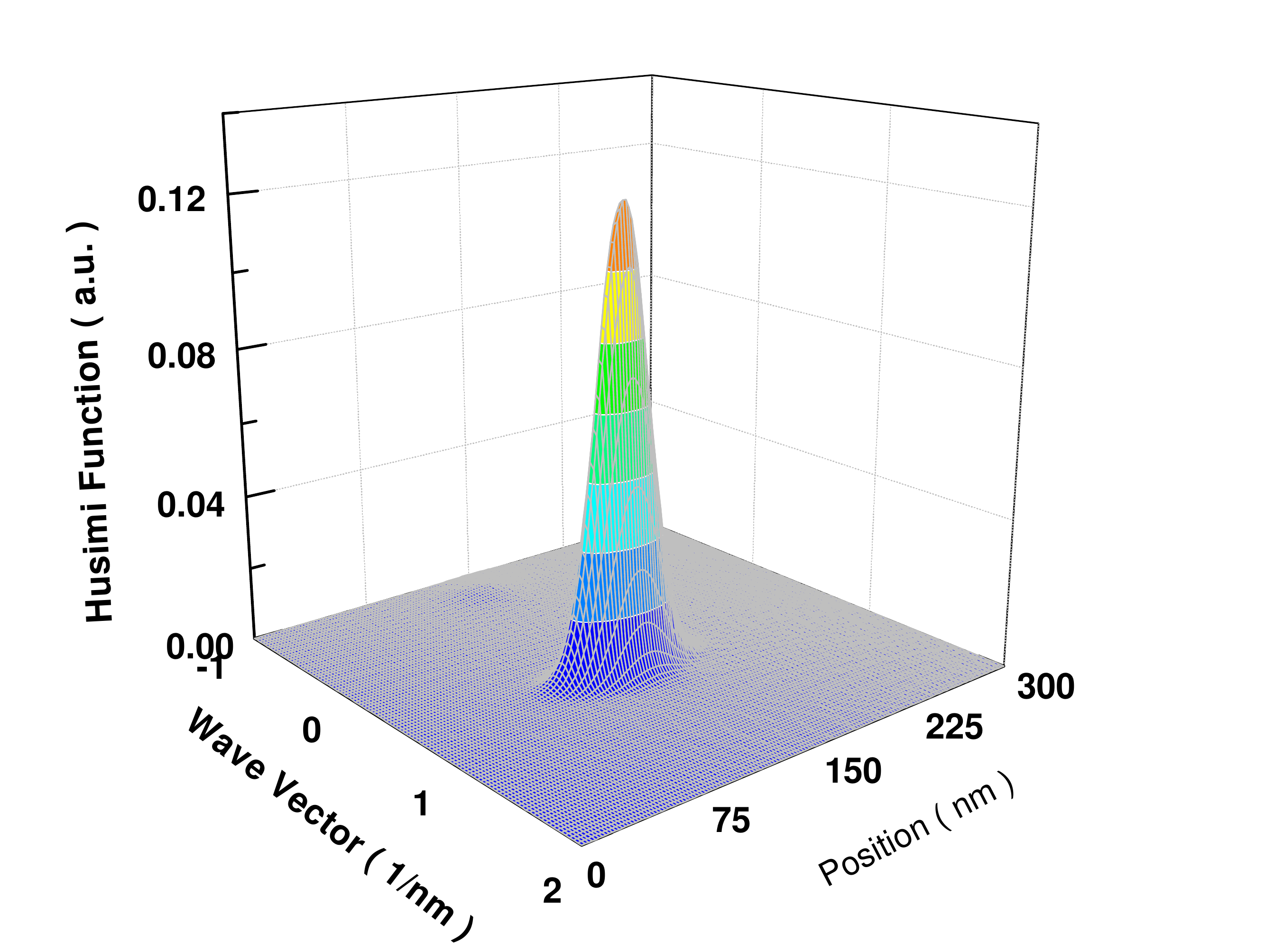}}
           \subfloat[Charge density for the three quantum phase space distributions at $t_1$.]{\label{ch1}
          \includegraphics[width=0.50\textwidth]{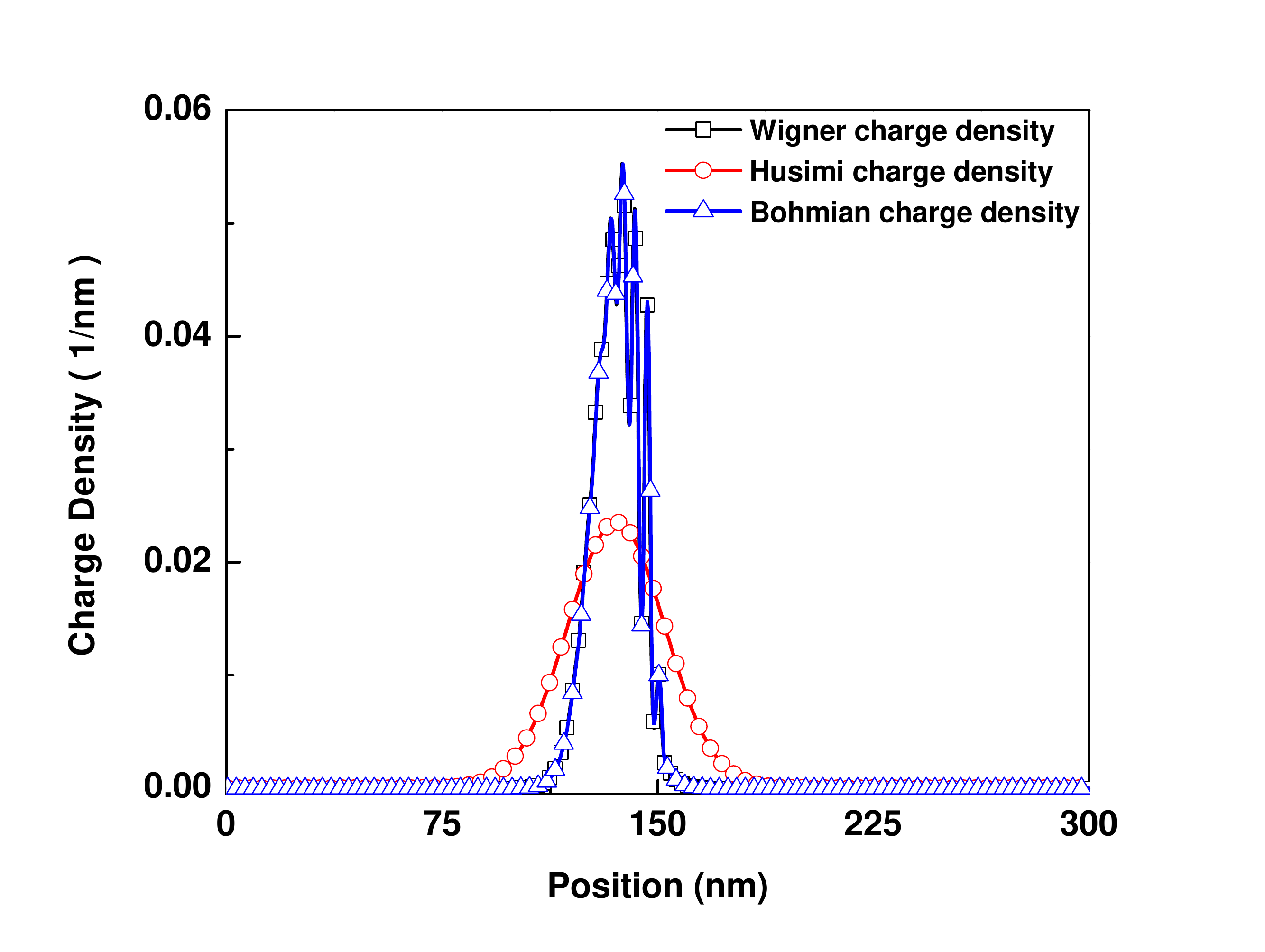}}\quad
          \subfloat[Bohmian distribution at $t_1$.]{\label{bf1}
          \includegraphics[width=0.50\textwidth]{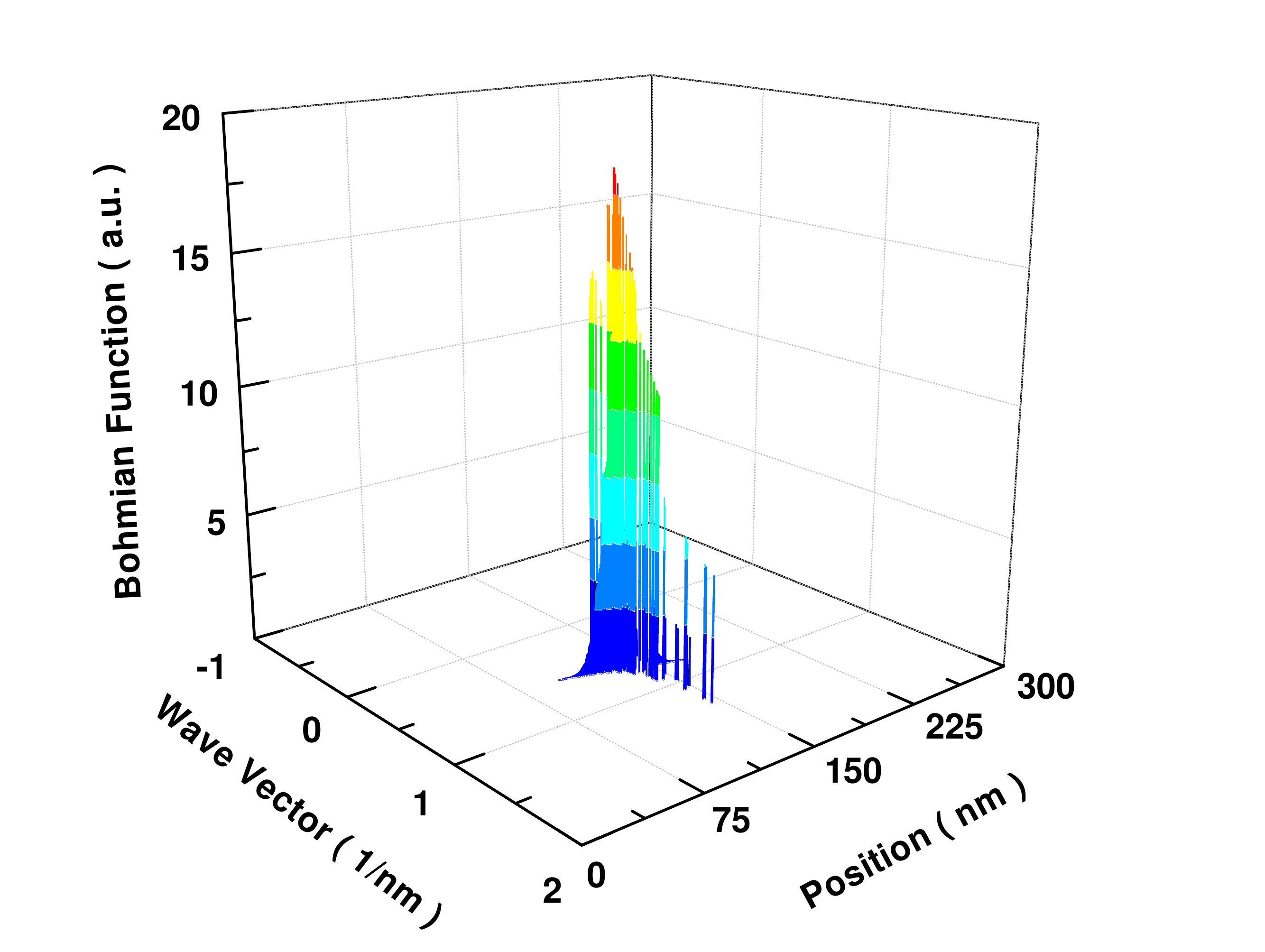}}
          \subfloat[Current density for the three quantum phase space distributions at $t_1$.]{\label{cu1}
          \includegraphics[width=0.50\textwidth]{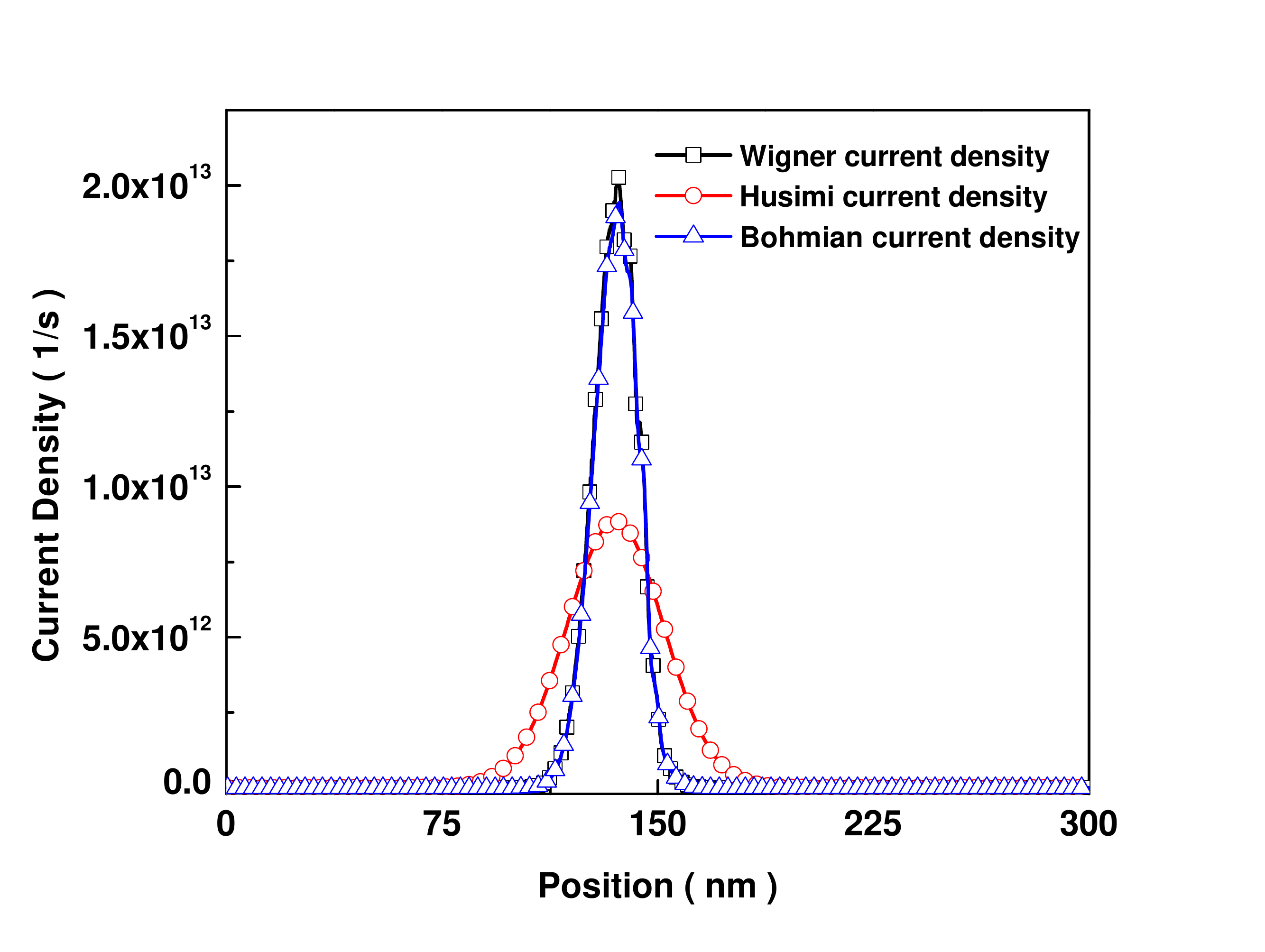}}
     \caption{Simulation of the (a) Wigner distribution, (c) Husimi distribution and (e) Bohmian distribution at the time $t_1$. (b) simulation of the wave packet impinging on a tunneling barrier with the same parameters as in Fig. \ref{fig1}. (d) and (f) are the charge density and current density for the three phase space distributions, respectively.}
  \label{fig2}
 \end{minipage}
\end{figure}

\begin{figure}[h]
\begin{minipage}{18cm}
  \centering
      \subfloat[Wigner distribution at $t_2$.]{\label{wf2}
          \includegraphics[width=0.50\textwidth]{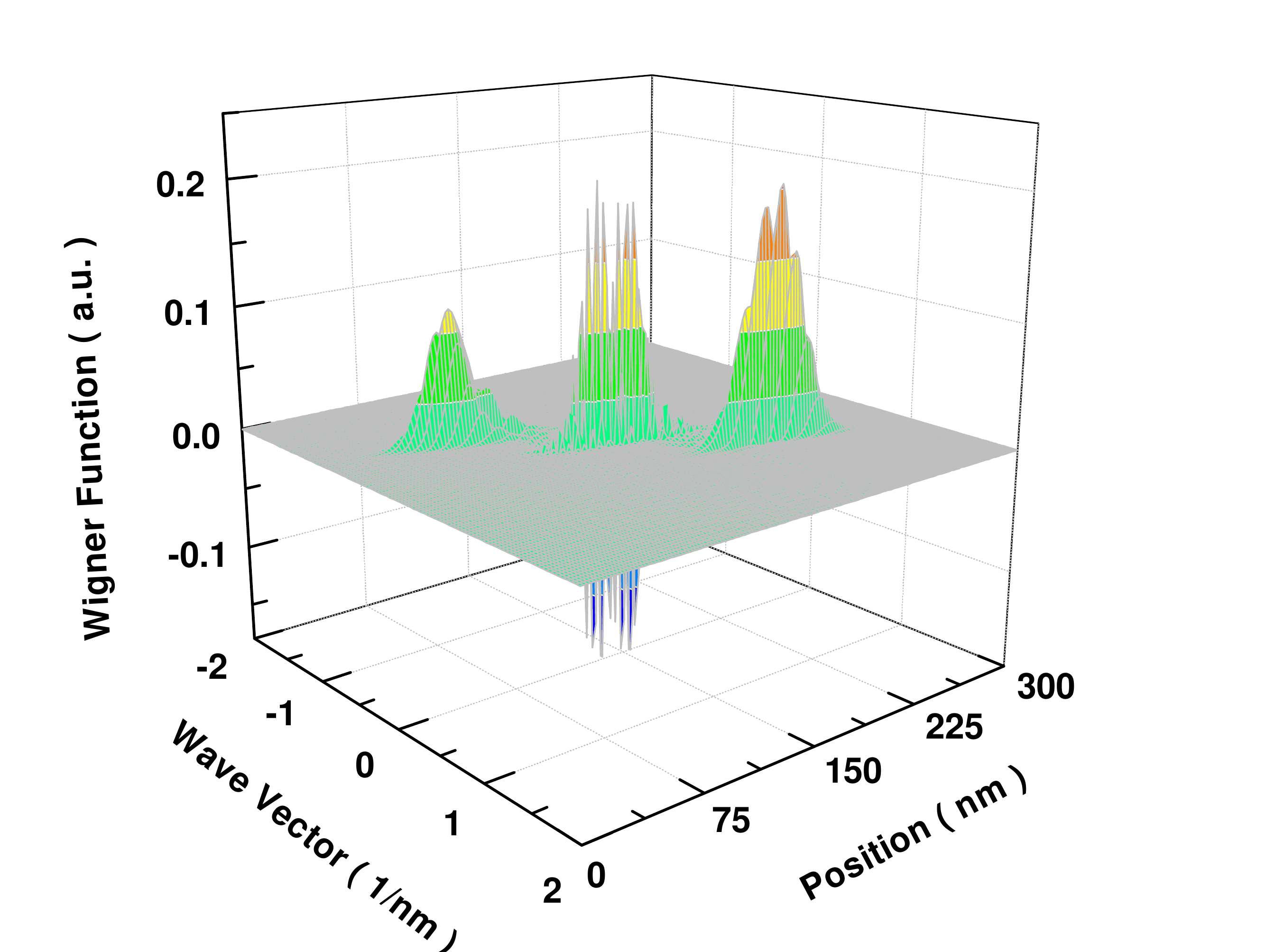}}
           \subfloat[Wave packet impinging on a tunneling barrier at $t_2$.]{\label{wp2}
          \includegraphics[width=0.50\textwidth]{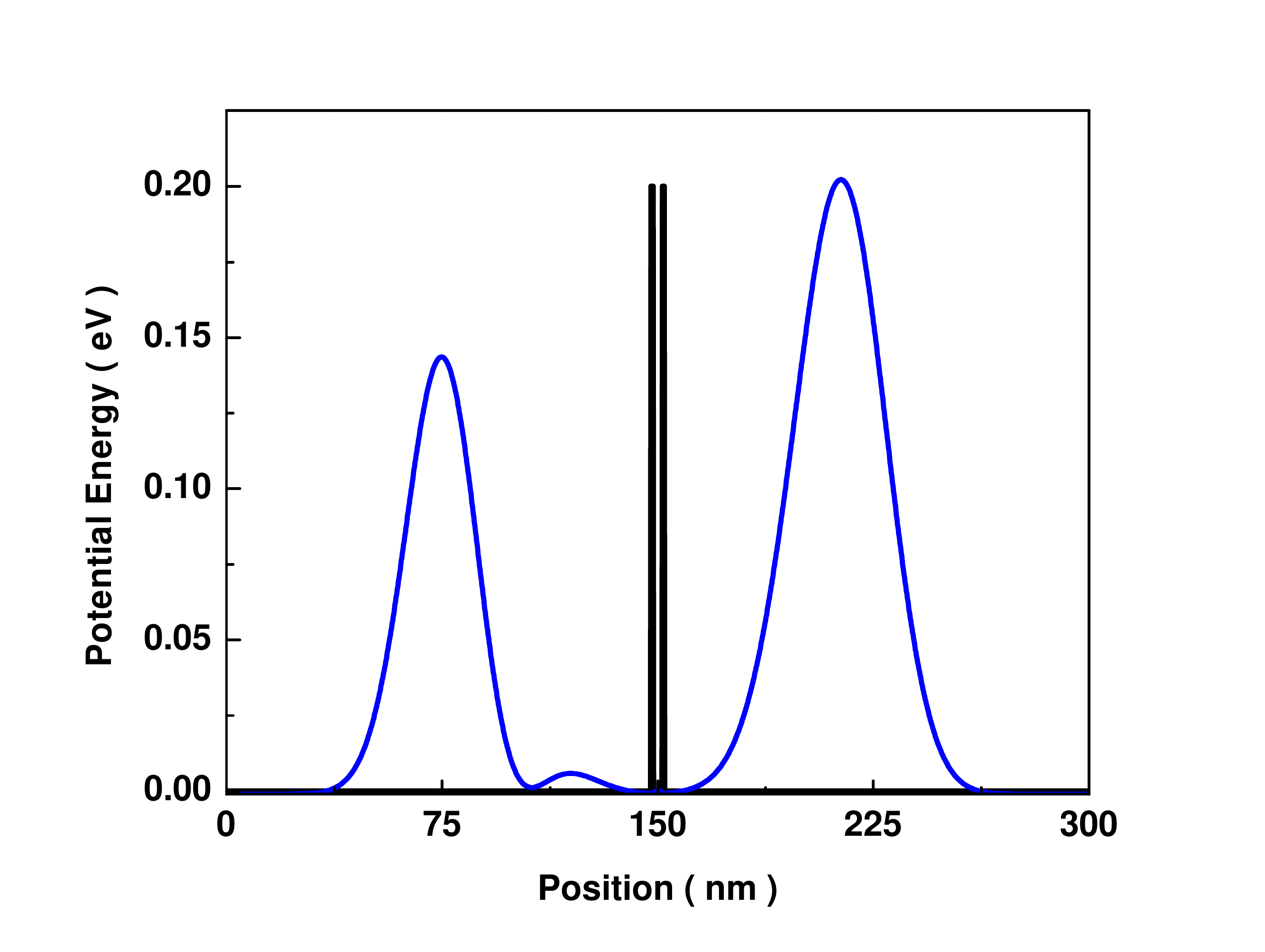}}\quad
          \subfloat[Husimi distribution at $t_2$.]{\label{hf2}
          \includegraphics[width=0.50\textwidth]{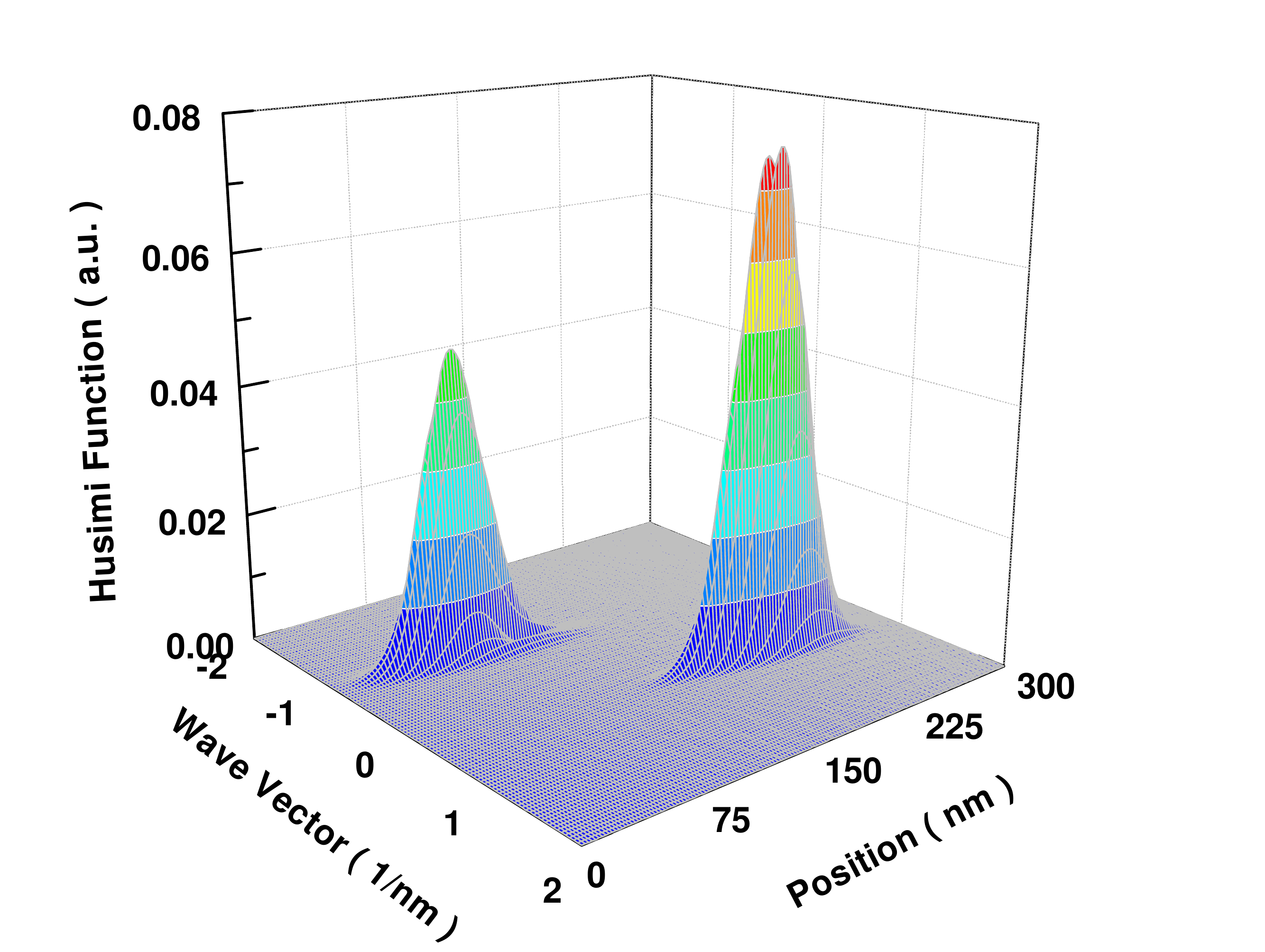}}
           \subfloat[Charge density for the three quantum phase space distributions at $t_2$.]{\label{ch2}
          \includegraphics[width=0.50\textwidth]{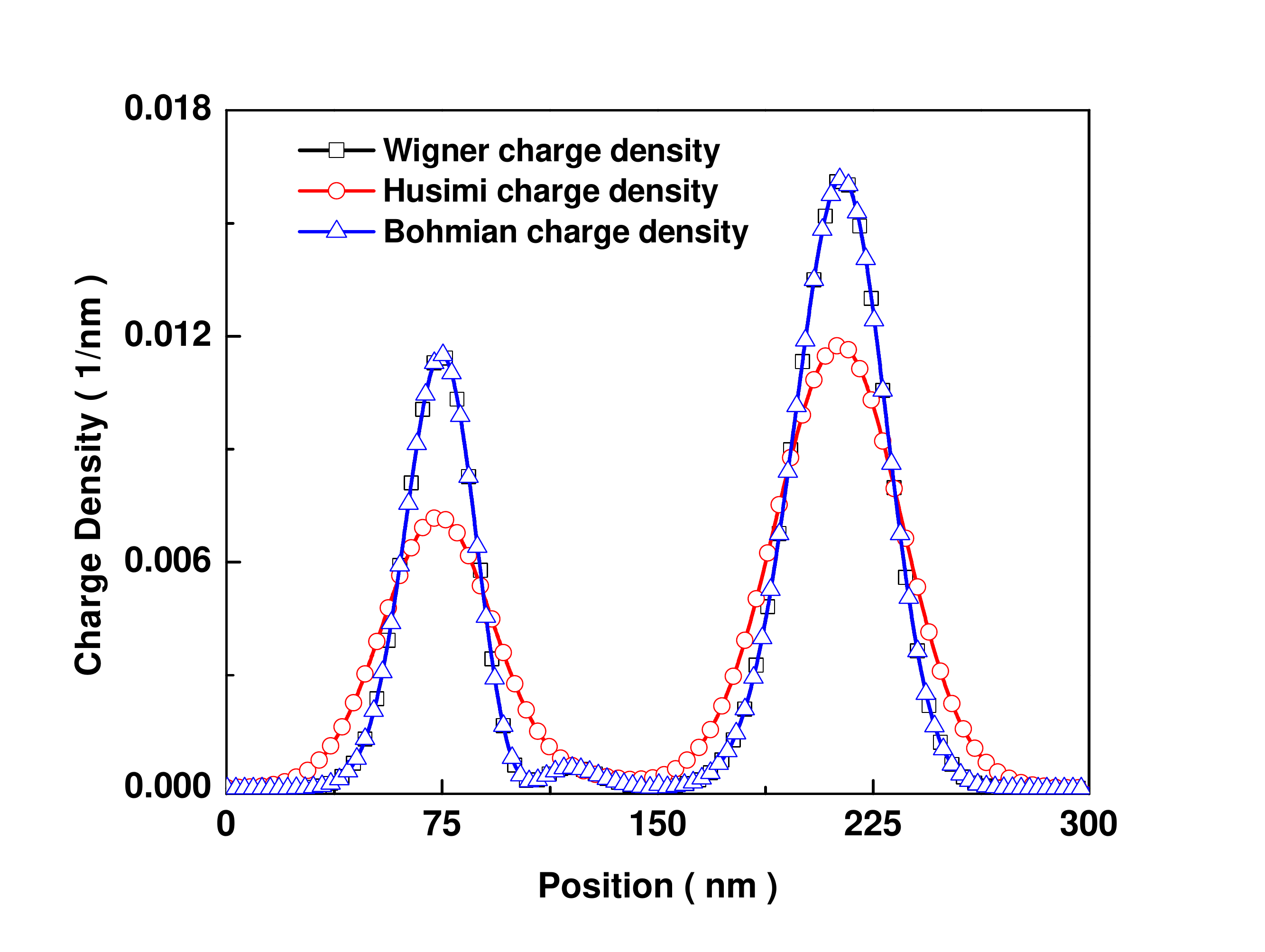}}\quad
          \subfloat[Bohmian distribution at $t_2$.]{\label{bf2}
          \includegraphics[width=0.50\textwidth]{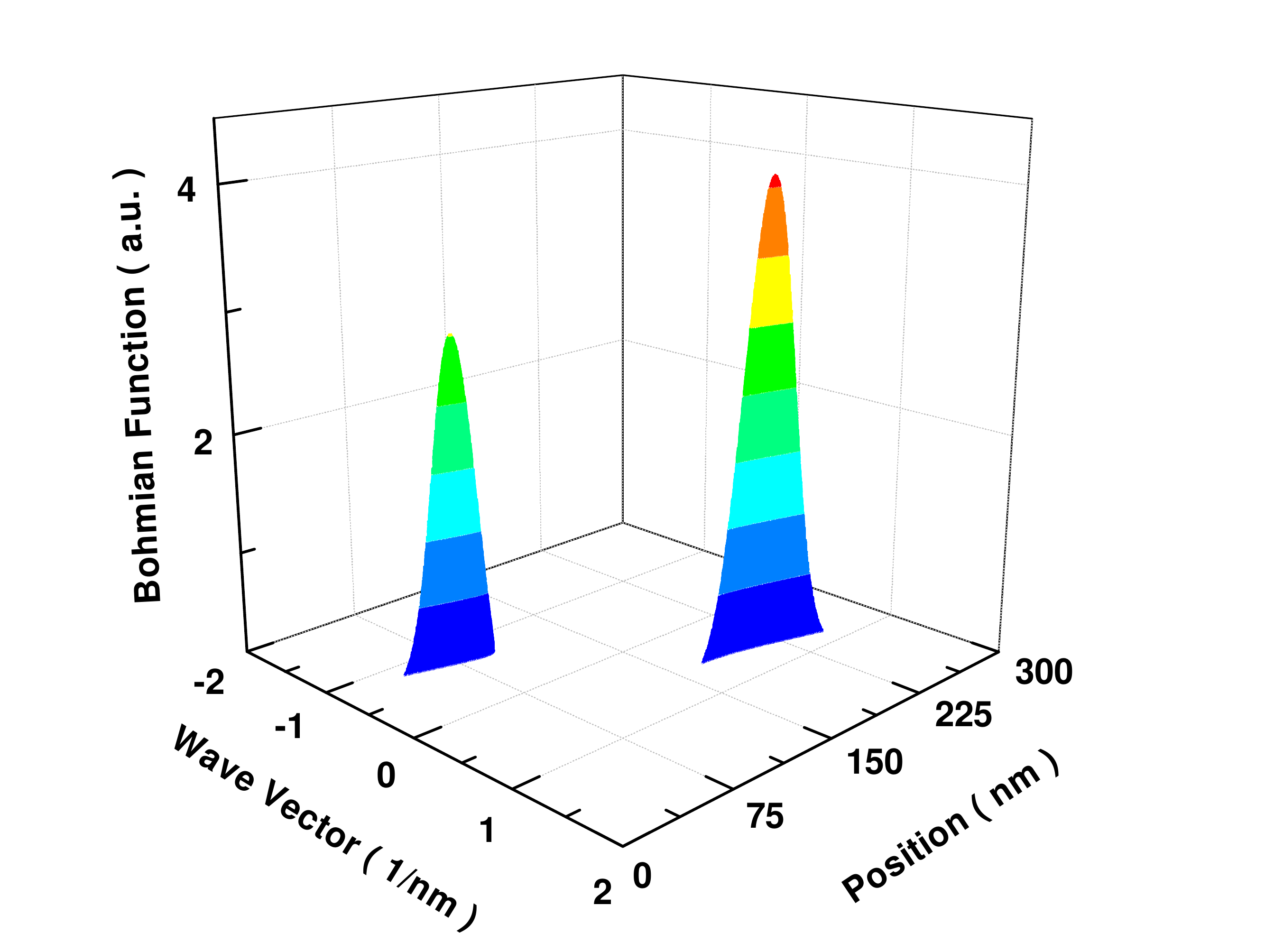}}
          \subfloat[Current density for the three quantum phase space distributions at $t_2$.]{\label{cu2}
          \includegraphics[width=0.50\textwidth]{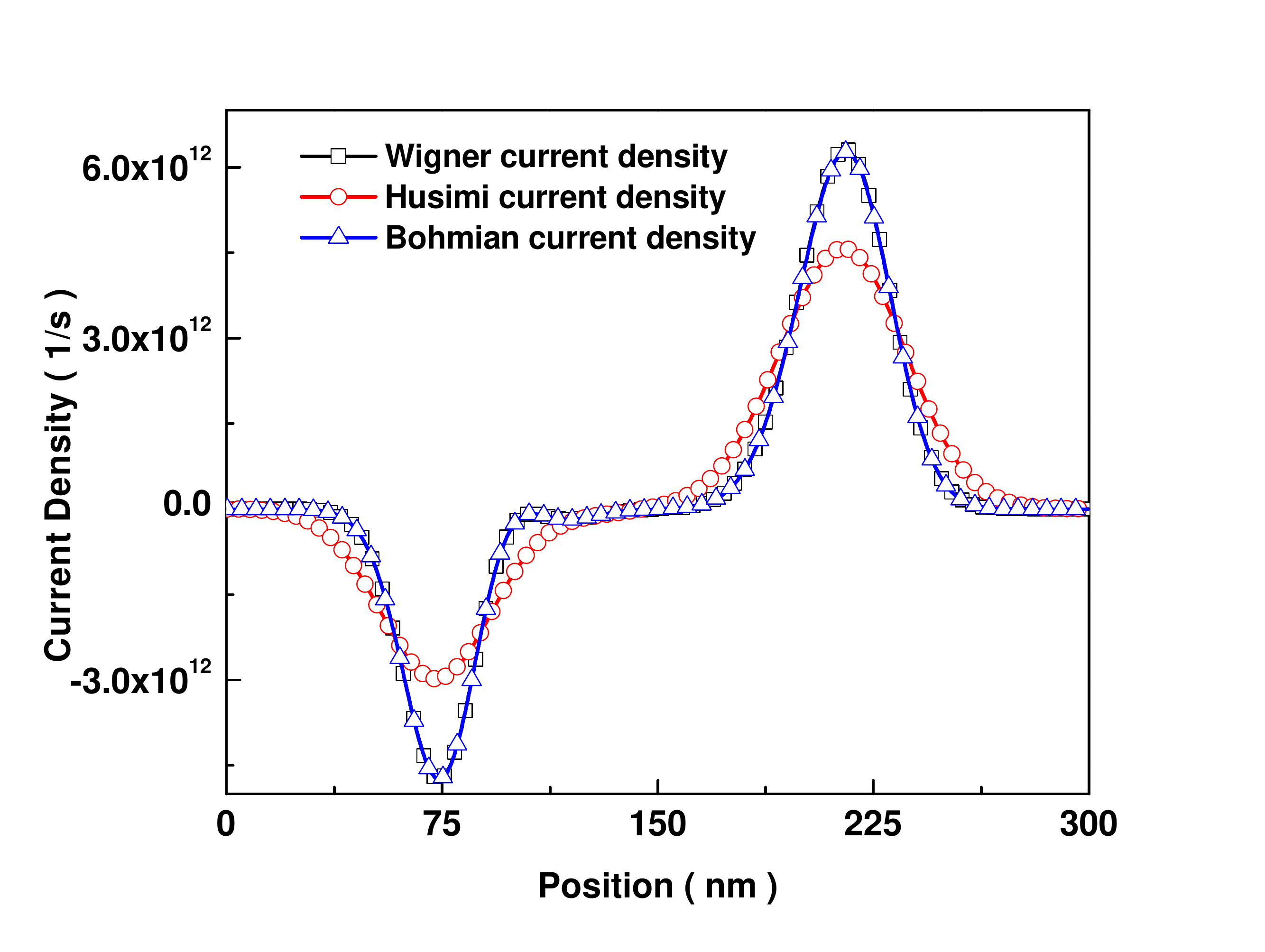}}
     \caption{Simulation of the (a) Wigner distribution, (c) Husimi distribution and (e) Bohmian distribution at the time $t_2$. (b) simulation of the wave packet impinging on a tunneling barrier with the same parameters as in Fig. \ref{fig1}. (d) and (f) are the charge density and current density for the three phase space distributions, respectively.}
  \label{fig3}
 \end{minipage}
\end{figure}

The time evolution of the initial wave packet is computed by numerically solving the Schr\"odinger equation (\ref{Scho}). Then, we compute the three quantum phase space distributions at three different times corresponding to the initial time $t_0=0\:ps$,  the time $t_1=0.09\:ps$ when the wave packet is interacting with the barrier and the time $t_2=0.3\:ps$ when the interaction is nearly finished and the initial wave packet is clearly split into a transmitted and a reflected components. The information corresponding to these three times are plotted in Figs. \ref{fig1}, \ref{fig2} and \ref{fig3}, respectively. 

Let us start by comparing the evolutions of the Wigner, Husimi and Bohmian distributions in Figs. \ref{fig1}-\ref{fig3}. It is clearly seen that the Bohmian and Husimi distributions have non-negative values everywhere at any time, satisfying clearly the first probability axiom (\ref{cond1}). At the initial time, the Wigner distribution is also non-negative, however, in later times at $t_1$ and $t_2$, negative values appears in some regions of the phase space. We will further discuss such unphysical feature and
\color{black} their consequences later. We also want to emphasize that the Bohmian distribution has only one value of the velocity at each position (this is just a consequence that the wave function can take only a single-value at each position.). In fact, a realistic example in quantum transport must deal with open systems. Then, the pure state has to be substituted by a mixed state (a sum of conditional wave functions in the Bohmian language) and the Bohmian distribution will provide a distribution of velocities (each conditional wave function will have its own velocity) at each position in a very natural way.  We just avoid the consideration of mixed states to simplify the present discussion. 

Next, we compare the charge and current densities calculated using \eref{marg1} and \eref{current} for the three quantum distributions. Let us emphasize again that, as discussed in \sref{sec2}, the values obtained from the Wigner and Bohmian distributions are always exactly equal. However, the values of \eref{Q_H} and \eref{J_H} for the Husimi distribution does not provide the correct charge and current densities obtained from the wave function. We clearly see that the modulus squared of the wave packet (blue lines) in Fig.\ref{wp0}, Fig.\ref{wp1} and Fig.\ref{wp2} are equivalent to the charge density of the Wigner and Bohmian distributions, but not to the Husimi one. 

After confirming, from the numerical simulations, the main features that we expect from the distributions (i.e. the negative values of the Wigner distribution, the mistaken results for the charge and current densities for the Husimi distribution and the success in both aspects of the Bohmian distribution) we further discuss an important undesired characteristic of the Wigner distribution. After the interaction with the double barrier, say at the time $t_2$, the initial wave packets $\psi(x,t)$ splits into a reflected $\psi_R(x,t)$ part  and a transmitted $\psi_T(x,t)$  part.
\clearpage

\clearpage

\clearpage

\clearpage
As seen in the plot Fig.\ref{wp2} of the modulus squared of the wave function, the reflected part and transmitted part are located at both sides of the barrier (not inside). 

Identically, in the Wigner distribution in Fig.\ref{wf2} we easily recognize the spatial locations of $F_W(x,p)$ from the reflected $\psi_R(x,t)$  and transmitted $\psi_T(x,t)$  parts. However, in addition, there are large non-zero values (negative and positive, because when integrating in this region of the phase space the result must be zero) of $F_W(x,p)$ in the middle of the barrier, $x=150\;nm$, at places where no probability presence of the electron is supposed to be according to Fig.\ref{wp2}.  The mathematical  reason of such spurious result is clear from the Wigner-Weyl transform. At time $t_2$, the spatially separated reflected $\psi_R(x,t)$  and transmitted $\psi_T(x,t)$ parts of the wave function $\psi(x,t)$ have to be spatially displaced a distance $y/2$ to compute $F_W(x,p)$ from  \eref{wigner-psi}. In particular, for the value of $y=2d$, we get  $\psi_R(x+d,t)$  and $\psi_T(x+d,t)$, plus $\psi^*_R(x-d,t)$  and $\psi^*_T(x-d,t)$. Finally, according to \eref{wigner-psi}, the $F_W(x,p)$ at time $t_2$ at position $x_C=150\;nm$ for any value of the momentum $p$ is: 

\begin{eqnarray}
F_W(x_C,p) \varpropto \frac{1}{h}  \psi_R(x_C+d) \psi_T^{*}(x_C-d)e^{i\frac{2 p d}{\hbar}} .
\label{error}
\end{eqnarray}
since $d$ is the distance between barrier and the reflected part, which we consider equal to the distance between the transmitted part of the barrier,  the product $\psi_R(x_C+d) \psi_T^{*}(x_C-d)$ in (\ref{error}) is different from zero at the barrier region $x_C$.

In order to clarify the consequences of this unphysical feature described in \eref{error} (we know from the wave function evolution that there are no electron probability there!), we remind that by integrating the positive and negative values of $F_W(x,p)$ around $x_C$ we will reproduce correctly the charge density at this point. 

There is no charge density at position $x_C=150\:nm$ in Fig.\ref{ch2}. However, if one tries to gives a physical meaning to $F_W(x,p)$ at these points  in Fig.\ref{wf2} (as a \emph{true} physical  probability distribution of the electron at the phase space), one must be very careful. For example, let us imagine that we introduce an \emph{ad-hoc} scattering term (due to impurities, for example) in the quantum equation of motion of the Wigner function. If such \emph{ad-hoc} term is introduced as a transition from an old phase space  point $\{x_C,p\}$ towards a new point $\{x_C,p\rq{}\}$ through the Fermi golden rule probability $S_{p,p'}$, we are moving electrons from places (for example $x_C=150\:nm$) where there are no electrons. The mistake in the transition of probability from regions without electrons is because we introduce the scattering mechanism by hand as an extra \emph{ad-hoc} term in the quantum equation of motion of the Wigner function. Obviously, this spurious effect will not be present if the scattering mechanism (with the impurity) is introduced directly in the Hamiltonian inside the quantum equation of motion. The mentioned undesired features will not occur within the Bohmian distribution, because as seen in Fig. \ref{bf2}, we only see non-zero (positive)  probabilities at locations where the electron may be reflected or transmitted, but not in other regions. 

\section{Conclusions}
\label{sec4}

In this work, we discuss the motivations and interests for constructing a \emph{true} (well-defined) phase space distribution for quantum systems. We have shown that, both, the Wigner and Husimi distributions (related to \emph{orthodox} or Cophengauhen quantum mechanics) do not satisfy this purpose completely and, in fact, they are called quasi-probability distributions.

The Wigner distribution, as seen in \eref{negative_wigner}, carries negative values that cannot be supported by a well-defined quantum probability distribution in phase space. Of course, when properly integrated, correct probability distributions are found out for the charge and current densities.  However,  the \emph{ad-hoc} manipulation of the Wigner distribution function can lead to incorrect results (we saw for instance in \sref{sec3} the non-zero values at places where there is no electron and we discuss the spurious transitions between regions without electrons when introducing scattering by hand as an extra term in the quantum equation of motion of the Wigner distribution). The Husimi distribution (a smoothed version of the Wigner distribution), by construction, only  has positive values. However, we cannot obtain successfully the charge and current densities. Thus it is not a proper probability distribution for quantum electron transport. 

In addition, we have presented the Bohmian distribution (associated to Bohmian mechanics) that fulfills all requirements to be a \emph{true} (well-defined) probability distribution. It is a positive probability distribution that exactly reproduces the charge and current densities used in the development of quantum electron transport simulators. These good properties are just a consequence of the Bohmian theory, which allows well-defined momentum and position, simultaneously, for an electron. The summary of the characteristics of the three distributions are seen in Table \ref{table}. 

\begin{table}[H]
\centering
\begin{tabular}{|p{3cm} ||p{1cm} |p{1cm} |p{1cm}|}
\hline
 & $F_W$ & $F_H$ & $F_B$ \\
\hline \hline
Positive distribution & No & Yes & Yes \\
\hline
Get the exact $Q(x)$ & Yes & No & Yes \\
\hline
Get the exact $J(x)$ & Yes & No & Yes \\
\hline
\end{tabular}
\caption{Bohmian distribution is the only that satisfies all requirements. Wigner and Husimi distributions are quasi-probability distributions and do not satisfy all requirements.}
\label{table}
\end{table}

Finally, we would like to clarify whether or not the Wigner and the Bohmian distributions can be measured. As we stress in the expression (\ref{commutator}) of the introduction, it is not possible to measure the position and the momentum of an electron, simultaneously, in a single experiment. From this unquestionable result, the Copenhagen school renounces to an (ontological) definition of the momentum at one particular position (i.e. renouncing, in fact, to a proper definition of the phase space), while the Bohmian school did not renounce to it. Nowadays, with our technological advances in carefully measuring (i.e. interacting with) a quantum system, the Wigner and Bohmian phase space distributions are, in fact, associable with experiments in a laboratory. In both cases, the \lq\lq{}trick\rq\rq{} is repeating the experiment many times (not only one) and treating the data  of such a large ensemble of experiments in a proper way.  Regarding the Wigner function, it is \lq\lq{}measured\rq\rq{} using the so-called quantum tomography. For example, one can measure the electric field quadratures \cite{Schiller} in many different experiments. Using these data and with the help of some mathematical expressions, a phase space can be reconstructed through the Wigner function distribution (with its negative values).  See an equivalent work in Ref. \cite{Monroe}  However, one must take into account that this (ensemble) phase space is not the \emph{true} phase space of a wave function in a single experiment (which does not exist in the orthodox quantum mechanics).

Identically, Bohmian phase space distribution can be measured by an ensemble of repeated experiments, through the use of weak measurements \cite{aharonov,povm}. This kind of measurement does not perturb so much the wave function as a strong (or projector) measurement, but the output result has a large uncertainty \cite{povm,oriols2}. With a proper treatment (the so-called weak value) of the data of an ensemble of weak and subsequent strong measurements of the positions and velocities, the Bohmian distribution can be constructed. In this regard, let us notice that the velocities of photons at particular positions have been already measured in a double slit experiment \cite{science}. Again, we remark that this result comes from an ensemble of experiments. Thus, again, it is just an ensemble phase space. We cannot extract this information from one unique measurement. In any case, contrarily to the Copenhagen school, Bohmian mechanics allows a well-defined probability distribution in phase space (according to the rules (\ref{cond1})-(\ref{current})). To be honest, every formalism has pros and cons and Bohmian phase space distributions is still in its infancy. In any case, in the authors' opinion, such \emph{true} phase space distribution has an enormous potential to be developed and exploited, in general, for study any (non-relativistic) quantum system and, in particular, for quantum electron transport. Some results justifying the viability of this path can be found in the BITLLES simulator developed by Oriols et al. \cite{damiano,xavier1,xavier2}.

\begin{acknowledgements}
This work has been partially supported by the \lq\lq{}Ministerio de Ciencia e Innovaci\'{o}n\rq\rq{} through the Spanish Project TEC2012-31330 and by the Grant agreement no: 604391 of the Flagship initiative  \lq\lq{}Graphene-Based Revolutions in ICT and Beyond\rq\rq{}. Z. Z acknowledges financial support from the China Scholarship Council (CSC).
\end{acknowledgements}

\begin{appendices}

\section{Q(x) and J(x) derivations}
\label{wigner_derivations}

In this appendix, we will develop the calculus to obtain the charge and current densities (Eqs. (\ref{marg1}) and (\ref{current})) for each of the three analysed distributions. 

\subsection{Wigner distribution}

The charge distribution is straightforwadly found out:

\begin{eqnarray}
&&Q_W(x,p) = \frac{1}{h} \int dp\int  \psi(x+\frac{y}{2}) \psi^{*}(x-\frac{y}{2})e^{i\frac{py}{\hbar}} dy = \nonumber \\
&=& \int \delta(y)  \psi(x+\frac{y}{2}) \psi^{*}(x-\frac{y}{2})e^{i\frac{py}{\hbar}} dy = |\psi(x)|^2.
\end{eqnarray}

The computation of $J_W (x)$ needs a more detailed discussion:

\begin{eqnarray}
\label{momentum}
&&J_W (x)= \int p F_W(x,p) dp= \nonumber \\
&=& \frac{1}{h}\int p \int \psi(x+\frac{y}{2}) \psi^{*}(x-\frac{y}{2})e^{i\frac{py}{\hbar}} dy dp.
\end{eqnarray}

Using the chain rule for a function $F(x,y)$ derivable and zero-valued at $y \to \pm \infty$, we can use the following relation

\begin{eqnarray}
\int dy e^{-ipy/\hbar} \frac{\partial}{\partial y} F(x,y) = \frac{i}{\hbar}p \int dy e^{-ipy/\hbar} F(x,y)
\label{chain}
\end{eqnarray}

and using the polar form of the wave function ($\psi(x)=R(x)e^{iS(x)}$), then rewrite \eref{momentum} as:

\begin{eqnarray}
\label{eq-numerator-wigner}
&&J_W (x) = \frac{-i}{2\pi} \int \! dp \int \! dy e^{-ipy/\hbar}
\! \cdot \! \Big[ R(x\!+\!\frac{y}{2})R(x\!-\!\frac{y}{2})\! \cdot \nonumber \\
&\cdot & e^{\frac{i}{\hbar}\left[ S(x+\frac{y}{2})-S(x-\frac{y}{2})\right]} \frac{i}{\hbar}\Big( \frac{\partial S(x+\frac{y}{2})}{\partial y} -\frac{\partial S(x-\frac{y}{2})}{\partial y} \Big) \!\!+ \nonumber \\
&+& e^{\frac{i}{\hbar}\left[ S(x+\frac{y}{2})-S(x-\frac{y}{2})\right]} \frac{\partial}{\partial y} \Big( R(x+\frac{y}{2})R(x-\frac{y}{2}) \Big) \Big]. \;\;
\end{eqnarray}

To proceed, let us focus on the following term:

\begin{eqnarray}
&&\frac{\partial S(x+\frac{y}{2})}{\partial y} \!\!-\!\!\frac{\partial S(x\!\!-\!\!\frac{y}{2})}{\partial y}\!\! =  \lim_{t \to 0} \Big[ \frac{S(x+\frac{y+t}{2})\!-\! S(x+\frac{y}{2})}{t} \!\!- \nonumber \\
&-&\!\! \frac{S(x\!-\!\frac{y+t}{2})\!\!-\!\! S(x\!-\!\frac{y}{2})}{t}\Big] \!\!= \!\! \frac{1}{2} \lim_{t \to 0}\Big[ \frac{S(x\!+\!\frac{t}{2} +  \frac{y}{2})\!-\!S(x\!+\!\frac{y}{2})}{t/2} \!\!+ \nonumber \\
 &+&\!\! \frac{S(x\!\!-\!\!\frac{y}{2})\!\!-\!\! S(x\!-\!\frac{t}{2}-\frac{y}{2})}{t/2} \Big]\!\!=\!\!  \frac{1}{2} \left[ \frac{\partial S(\!x\!+\!\frac{y}{2})}{\partial x} \!+\! \frac{\partial S(x\!-\!\frac{y}{2})}{\partial x} \!\right]\!,
\label{aux1}
\end{eqnarray}

which can be used to rewrite \eref{eq-numerator-wigner} as

\begin{eqnarray}
&&J_W (x) \!\!=\!\! \int \!\! dy \delta(y) \! \Big[ R(x\!+\!\!\frac{y}{2})R(x\!\!-\!\!\frac{y}{2}) e^{\frac{i}{\hbar}\left[ S(x+\frac{y}{2})-S(x-\frac{y}{2})\right]}\cdot \nonumber \\
& \cdot & \!\! \frac{1}{2}\Big( \frac{\partial S(x+\frac{y}{2})}{\partial x} +\frac{\partial S(x-\frac{y}{2})}{\partial x} \Big) - i\hbar e^{\frac{i}{\hbar}\left[ S(x+\frac{y}{2})-S(x-\frac{y}{2})\right]} \cdot \nonumber \\
&\cdot & \frac{\partial}{\partial y} \Big( R(x+\frac{y}{2})R(x-\frac{y}{2}) \Big) \Big] =  R^{2}(x)\frac{\partial S(x)}{\partial x}.
\label{Jwigner}
\end{eqnarray}

\eref{Jwigner} is the expression for the current density distribution for the Wigner function.

In \eref{Jwigner}  we have used the following property:
\begin{eqnarray}
\label{delta}
\int dp e^{-ipy/\hbar} = 2\pi \hbar \delta(y),
\end{eqnarray}
and the fact that the second term within the integral in \eref{Jwigner} is zero:
\begin{eqnarray}
\label{eq-R0}
 &&\frac{\partial}{\partial y} \Big( R(x+\frac{y}{2})R(x-\frac{y}{2}) \Big) \Big|_{y=0} = \nonumber \\
 &=& \frac{1}{2}\left[  \frac{\partial R(x+\frac{y}{2})}{\partial x} R(x-\frac{y}{2}) -  R(x+\frac{y}{2})\frac{\partial R(x-\frac{y}{2})}{\partial x} \right] \Big|_{y=0} \!\!\!\!= \nonumber \\
 &=& 0.
 \label{aux2}
\end{eqnarray}

\subsection{Husimi distribution}

The derivation of the Husimi charge and current densities are quite similar to the derivation realized for the Wigner distribution, with the only difference that now, the Wigner distributions are smoothed with a Gaussian function. In first place, we derive the charge distribution:

\begin{eqnarray}
&&Q_H(x,p)\! = \! \frac{1}{\pi \hbar}\int \!\!\int \!\! \int \!\! \frac{1}{h} \int \!\! \psi( x'\!-\!\frac{y}{2})\psi^*(x'\!+\!\frac{y}{2}) e^{i\frac{p'y}{\hbar}}dy \cdot  \nonumber \\
&\cdot \!\! &\!e^{-\frac{(x-x')^2}{2s^2}}e^{-\frac{(p-p')^22s^2}{\hbar^2}}dx'dp'dp = \frac{1}{\pi \hbar}\frac{1}{\sqrt{(2\pi s^2)}}\cdot  \nonumber \\
&\cdot &\int \!\! \int \!\! \int \! \psi( x'\!-\!\frac{y}{2})\psi^*(x'\!+\! \frac{y}{2}) e^{i\frac{p'y}{\hbar}}dy e^{-\frac{(x-x')^2}{2s^2}}dx'dp'= \nonumber \\
&=&\!\frac{1}{\sqrt{(2\pi s^2)}}\int \!\! \int \!\! \psi( x'\!-\! \frac{y}{2})\psi^*(x'\!+\! \frac{y}{2}) \delta(y)dy e^{-\frac{(x-x')^2}{2s^2}}dx'= \nonumber \\
&=&\frac{1}{\sqrt{(2\pi s^2)}}\int  |\psi( x')|^2 e^{-\frac{(x-x')^2}{2s^2}}dx'.
\label{Husimi_charge}
\end{eqnarray}

After that, we derive the current density. For this, Eqs. (\ref{chain}), (\ref{aux1}), (\ref{delta}) and (\ref{aux2}) will be used again:

\begin{eqnarray}
\label{momentum_H}
&& J_H (x)\!\!= \!\!\frac{1}{\pi \hbar}\!\!\int \!\! \!\!\int \!\!\!\! \int \!\! p F_H(x',p')e^{-\frac{(x-x')^2}{2s^2}}e^{-\frac{(p-p')^2}{2\sigma_p^2}}dx'dp' dp\!= \nonumber \\
&=&\frac{1}{\pi \hbar}\int \!\!\!\! \int \!\!\!\! \int \!\! p \frac{1}{h} \int \!\!\psi( x'\!\!-\!\!\frac{y}{2})\psi^*(x'\!\!+\!\! \frac{y}{2}) e^{i\frac{p'y}{\hbar}}dy e^{-\frac{(x-x')^2}{2s^2}} \cdot  \nonumber \\ 
&\cdot &e^{-\frac{(p-p')^22s^2}{\hbar^2}}dx'dp' dp =\!\!\frac{1}{h}\frac{1}{\sqrt{(2\pi s^2)}}\int \!\! p'\psi( x'\!\!-\!\! \frac{y}{2})\cdot  \nonumber \\
&\cdot & \psi^*(x'+\frac{y}{2}) e^{i\frac{p'y}{\hbar}}dy e^{-\frac{(x-x')^2}{2s^2}}dx'dp'= \nonumber \\
&=&\frac{1}{\sqrt{(2\pi s^2)}}\int R^{2}(x')\frac{\partial S(x')}{\partial x'}e^{-\frac{(x-x')^2}{2s^2}}dx'
\end{eqnarray}

Therefore, we can clearly see that these results for the charge and current densities are different from the Wigner results.

\subsection{Bohmian distribution}

The charge distribution is the following:

\begin{eqnarray}
&&Q_B(x,t)\!=\!\int\!\!\lim_{N \to \infty}\frac{1}{N}\sum_{i=1}^N\!\delta(x\!-\!x_i(t))\delta(p\!-\!p_i(t))dp = \nonumber \\
&=& \lim_{N \to \infty}\frac{1}{N}\sum_{i=1}^N\delta(x-x_i(t))=|\psi(x,t)|^2.
\label{charge_bohmian}
\end{eqnarray}

The last equality in \eref{charge_bohmian} is due to the quantum equilibrium hypothesis, which states that the charge distribution at time $t$ is equal to the modulus squared of the wave function (for a more detailed discussion see \cite{Durr1992}). 

In order to calculate the current density, we need to express the wave function in the polar form (in the same way as done for the Wigner distribution) and we proceed in the following way:

\begin{eqnarray}
&&J_B(x,t)=\lim_{N \to \infty}\frac{1}{N}\sum_{i=1}^Np\delta(x-x_i(t))\delta(p-p_i(t))dp = \nonumber \\
&=& \lim_{N \to \infty}\frac{1}{N}\sum_{i=1}^Np_i\delta(x-x_i(t))=R^{2}(x)\frac{\partial S(x)}{\partial x}.
\label{charge_bohmian2}
\end{eqnarray}

Apart from using again the quantum equilibrium hypothesis, we have also used that the momentum in Bohmian mechanics in a single-particle case is \cite{Durr1992}:

\begin{eqnarray}
p(x,t) =  \text{Im} \frac{\nabla \psi}{\psi}\equiv \frac{\partial S(x)}{\partial x}.
\label{momentum_bohmian_particle}
\end{eqnarray}

Let us notice that \eref{momentum_bohmian_particle} is just \eref{vel} multiplied by the electron mass $m$.

\section{Negative values of the Wigner distribution}
\label{neg_wig}

If we manipulate the Wigner distribution (applying a change of variable), we can see that it can be thought as a correlation function:

\begin{eqnarray}
&&F_W(x,p) = \frac{1}{h} \int dp\int  \psi(x+\frac{y}{2}) \psi^{*}(x-\frac{y}{2})e^{i\frac{py}{\hbar}} dy = \nonumber \\
&=& 2\int \psi(r) \psi^{*}(2x-r)e^{i\frac{py}{\hbar}}e^{i\frac{2p(r-2x)}{\hbar}} dr = 2\int \psi(r)e^{i\frac{pr}{\hbar}} \cdot \nonumber \\
&\cdot &\!\!\! \big(\psi(2x-r)  e^{-i\frac{2p(r-2x)}{\hbar}}\big)^{*} dr\! = \! 2\int \!\! \phi(r) \phi^{*}(2x-r) dr.
\label{correlation_w}
\end{eqnarray}

The change of variable is the following: $x+\frac{y}{2}=r$. In addition, we have defined $\phi(r)=\psi(r)e^{i\frac{pr}{\hbar}}$. For simplicity we do not indicate the dependence on $p$.

With these considerations, we can prove that the Wigner function is real and can be negative. We compute the integral of a modulus of a function $|\phi(r) + \phi(2x-r)|^2$ which is obviously positive:

\begin{eqnarray}
&&\int\big(\phi(r) + \phi(2x-r)\big) \cdot \big(\phi(r)^{*} + \phi^{*}(2x-r)\big)dr = \nonumber \\
&=&\!\int\!\!|\phi(r) + \phi(2x-r)|^2dr \!\!=\!\!\int\!\Big(|\phi(r)|^2+|\phi(2x-r)|^2 +\nonumber \\
&+& \phi(r)\phi^{*}(2x-r)+\phi(r)^{*}\phi(2x-r)\Big)dr\geq 0.
\label{negative_wigner0}
\end{eqnarray}

Next, we observe that the last two terms in the right hand side are identical and, by using \eref{correlation_w}, equal to the Wigner distribution function $F_W(x,p)$. in order to see that both terms are identical, we make a change of variable $r'=2x-r$:

\begin{eqnarray}
&&\int\phi(r)^{*}\phi(2x-r)dr=-\!\!\int^{-\infty}_\infty\phi(r')^{*}\phi(2x-r')dr'= \nonumber \\
&=&\int\phi(r')^{*}\phi(2x-r')dr'.
\label{negative_wigner1}
\end{eqnarray}

Therefore, the sum of both terms are exactly equal to $F_W(x,p)$ in \eref{correlation_w}.

\begin{eqnarray}
&&\int|\phi(r) + \phi(2x-r)|^2dr =  \int|\phi(r)|^2dr+ \nonumber \\
&+&\int|\phi(2x-r)|^2dr + 2\int\phi(r)\phi^{*}(2x-r)dr\geq 0.
\label{negative_wigner}
\end{eqnarray}

that we rewrite as: 

\begin{eqnarray} 
&&F_W(x,p)=\int|\phi(r) + \phi(2x-r)|^2dr \nonumber \\ &-& \int|\phi(r)|^2dr -\int|\phi(2x-r)|^2dr 
\label{negative_wigner2} 
\end{eqnarray} 

Therefore, the value of the Wigner function in \eref{negative_wigner2} must be real and it can clearly take negative values when the first integral is smaller than the sum of the other two (for example, when $\phi(r)=-\phi(2x-r)$). In this way, it is proved that the Wigner function is a quasi-probability distribution, but not a true one.

\end{appendices}


\begin{thebibliography}{99}

\bibitem{broglie} 
de Broglie, L.: Remarques sur la nouvelle m\'ecanique ondulatoire. C. R. Acad. Sci. \textbf{183}, 272 (1926) 

\bibitem{omi.Planck-BlackBody} 
Planck, M.: On the Law of Distribution of Energy in the Normal Spectrum. Annalen der Physik, \textbf{4},   553 (1901)

\bibitem{omi.Einstein-Photoelectric} 
Einstein, A.: �ber einen die Erzeugung und Verwandlung des Lichtes betreffenden heuristischen Gesichtspunkt. Annalen der Physik, \textbf{17}, 132 (1905)

\bibitem{omi.bohr} 
Bohr, N.: On the Constitution of Atoms and Molecules, Part I. Philosophical Magazine, \textbf{26}, 1 (1913)\\
Bohr, N.: On the Constitution of Atoms and Molecules, Part II Systems Containing Only a Single Nucleus. Philosophical Magazine, \textbf{26}, 476 (1913)\\
Bohr, N.: On the Constitution of Atoms and Molecules, Part III Systems containing several nuclei. Philosophical Magazine, \textbf{26}, 857 (1913)

\bibitem{omi.dB_AnnPhys} 
de Broglie, L.: Recherches sur la th\'{e}orie des quantas. Ann. de Physique, \textbf{3}, 22 (1925)

\bibitem{omi.debroglie1927b} 
de Broglie, L.: La m\'{e}canique ondulatorie et la structure atomique de la mati\`ere et du rayonnement. Journal de Physique et du Radium, \textbf{8}, 225 (1927)

\bibitem{omi.bohm1952a} 
Bohm, D.: A suggested interpretation of the quantum theory in terms of ``hidden'' variables  I. Phys. Rev. \textbf{85}, 166, (1952)

\bibitem{omi.bohm1952b} 
Bohm, D.: A suggested interpretation of the quantum theory in terms of ``hidden'' variables  II. Phys. Rev. \textbf{85}, 180, (1952)

\bibitem{omi.Bohm1993}
Bohm, D., Hiley B. J.: The Undivided Universe: An Ontological Interpretation of Quantum Theory. Routledge \& Kegan Paul, London 1993

\bibitem{OriolsBook} 
Oriols, X., Mompart J.: Applied Bohmian Mechanics: From Nanoscale Systems to Cosmology. Pan Stanford Publishing, Singapore (2011)

\bibitem{omi.Born1926} 
Born, M.: Zur Quantenmechanik der Sto�vorg\"{a}nge. Zeitschrift f\"{u}r Physik, \textbf{37}, 863 (1926)

\bibitem{omi.Heisenber1925} 
Heisenberg, W.: \"{U}ber quantentheoretishe Umdeutung kinematisher und mechanischer Beziehungen. Zeitschrift f\"{u}r Physik, \textbf{33}, 879 (1925)

\bibitem{weyl}
Weyl, H.: Quantenmechanik und gruppentheorie. Z. Phys. A \textbf{46}, 1 (1927)

\bibitem{omi.wigner} 
 Wigner, E. P.: On the Quantum Correction For Thermodynamic Equilibrium. Phys. Rev., \textbf{40}, 749 (1932)
 
\bibitem{groenewold}
Groenewold,H. J.: On the principles of elementary quantum mechanics. Physica \textbf{12}, 405 (1946)

\bibitem{moyal}
Moyal, J. E.: Quantum mechanics as a statistical theory. Proc. Cambridge Philos. Soc. \textbf{45}, 99 (1949)

\bibitem{bayen1}
Bayen, F., Flato, M., Fronsdal, C., Lichnerowicz, A., Sternheimer, D.: Deformation theory and quantization. I. Deformations of symplectic structures. Ann. Phys. (N.Y.) \textbf{111}, 61 (1978)

\bibitem{bayen2}
Bayen, F., Flato, M., Fronsdal, C., Lichnerowicz, A., Sternheimer, D.: Deformation theory and quantization. II. Physical applications. Ann. Phys. (N.Y.) \textbf{111}, 111 (1978)

\bibitem{baker}
Baker, G. A.: Formulation of quantum mechanics based on the quasi-probability distribution induced on phase space. Phys. Rev. \textbf{109}, 2198 (1958)

\bibitem{Zachos}
Zachos, C. K., Fairlie, D. B., Curtright, T. L.: Quantum Mechanics in Phase Space. World Scientific Pub Co, Singapore (2005)

\bibitem{Dirac}
Dirac, P. A. M.: The Principles of Quantum Mechanics. Oxford University Press, Oxford (1991)

\bibitem{Monroe}
Leibfried, D., Pfau, T., Monroe, C.: Reconstructing Quantum States of Atom Motion. Physics Today, April (1998)

\bibitem{feynman}
Feynman, R. P.: Negative Probability in Quantum Implications. Essays in Honour of David Bohm, Routledge \& Kegan Paul Ltd, London \& New York (1987)

\bibitem{review}
Muckenheim, W., Ludwig, G., Dewdney, C., Holland, P. R., Kyprianidis, A., Vigier, J. P., Cufaro Petroni, N., Bartlett, M. S., Jaynes, E. T.: A review of extended probabilities. Phys. Rep. \textbf{133}, 337 (1986)

\bibitem{omi.husimi}  
Husimi, K.: Some formal properties of the density matrix. Proceedings of the Physico-Mathematical Society of Japan, \textbf{22}, 264 (1940)

\bibitem{kerry} 
Shifren, L., Kerry, D. K.: A Wigner function based ensemble Monte Carlo Approach for accurate incorporation of quantum effects in device simulation. J. Comput. Electron. \textbf{1}, 55 (2002)

\bibitem{frensley} 
Frensley, W. R.: Wigner-function of a resonant-tunneling semiconductor device. Phys. Rev. B \textbf{36}, 1570 (1988)

\bibitem{sellier}
Sellier, J. M., Nedjalkov, M., Dimov, I.: An introduction to applied quantum mechanics in the Wigner Monte Carlo formalism. Physiscs Reports \textbf{577}, 1 (2015)

\bibitem{weinbub}
Ellinghaus, P., Weinbub, J., Nedjalkov, M., Selberherr, S., Dimov, I.: Distributed-Memory Parallelization of the Wigner Monte Carlo Method using spatial domain decomposition. J. Comput. Electron. \textbf{14}, 151 (2015)

\bibitem{Ballentine}
Ballentine, L. E.: Quantum mechanics: a modern development, chapter 15. World Scientific, Singapore (1988)

\bibitem{Schiller}
Lvovsky, A. I., Hansen, H., Aichele, T., Benson, O., Mlynek, J., Schiller, S.: Quantum state reconstruction of the single-photon fock state. Phys. Rev. Lett. \textbf{87} 050402 (2001)

\bibitem{aharonov}
Aharonov, Y., Albert, D. Z., Vaidman, L.: How the result of a measurement of a component of the spin of a spin-1/2 can turn out to be 100. Phys. Rev. Lett. \textbf{60}, 1351 (1988)

\bibitem{povm}
Kofman, A. G., Ashhab, S., Nori, F.: Nonperturbative theory of weak pre- and post-selected measurements. Physics reports \textbf{520}, 43 (2012)

\bibitem{oriols2}
Traversa, F. L., Albareda, G., Di Ventra., M, Oriols, X.: Robust weak-measurement protocol for Bohmian velocities. Phys. Rev. \textbf{A}, 052124 (2013)

\bibitem{science}
Kocsis, S., Braverman, B., Ravets, S., Stevens, M. J., Mirin, R. P., Shalm, L. K., Steinberg, A. M.: Observing the average trajectories of single photons in a two-slit interferometer. Science \textbf{3}, 1170 (2011)

\bibitem{damiano}
Marian, D., Colom\'es, E., Zhen, Z., Oriols, X.: Quantum Noise from a Bohmian perspective: fundamental understanding and practical computation. J. Comput. Electron. \textbf{14}, 114 (2015)

\bibitem{xavier1}
Oriols, X.: Quantum trajectory approach to time dependent transport in mesoscopic systems with electron-electron interactions. Phys. Rev. Lett. \textbf{98}, 066803 (2007)

\bibitem{xavier2}
Traversa, F.L., et al.: Time-Dependent Many-Particle Simulation for Resonant Tunneling Diodes: Interpretation of an Analytical Small-Signal Equivalent Circuit. IEEE Trans. Elect. Dev. \textbf{58}, 2104-2112 (2011)

\bibitem{Durr1992} 
D\"urr, D., Goldstein, D., Zangh\`i, N.: Quantum equilibrium and the origin of absolute uncertainty. J. Stat. Phys. \textbf{67}, 843 (1992)

\end{thebibliography}
\end{document}